
\input harvmac.tex



\let\includefigures=\iftrue
\newfam\black
\includefigures
\input epsf
\def\figin{\epsfcheck\figin}\def\figins{\epsfcheck\figins}
\def\epsfcheck{\ifx\epsfbox\UnDeFiNeD
\message{(NO epsf.tex, FIGURES WILL BE IGNORED)}
\gdef\figin##1{\vskip2in}\gdef\figins##1{\hskip.5in}
\else\message{(FIGURES WILL BE INCLUDED)}%
\gdef\figin##1{##1}\gdef\figins##1{##1}\fi}
\def\DefWarn#1{}
\def\figinsert{\goodbreak\midinsert}
\def\ifig#1#2#3{\DefWarn#1\xdef#1{fig.~\the\figno}
\writedef{#1\leftbracket fig.\noexpand~\the\figno}%
\figinsert\figin{\centerline{#3}}\medskip\centerline{\vbox{\baselineskip12pt \advance\hsize by
-1truein\noindent\footnotefont{\bf Fig.~\the\figno:} #2}}
\bigskip\endinsert\global\advance\figno by1}
\else
\def\ifig#1#2#3{\xdef#1{fig.~\the\figno}
\writedef{#1\leftbracket fig.\noexpand~\the\figno}%
\global\advance\figno by1} \fi

\font\cmss=cmss10 \font\cmsss=cmss10 at 7pt

\def\IB{\relax\hbox{$\inbar\kern-.3em{\rm B}$}}
\def\IC{\relax\hbox{$\inbar\kern-.3em{\rm C}$}}
\def\IQ{\relax\hbox{$\inbar\kern-.3em{\rm Q}$}}
\def\ID{\relax\hbox{$\inbar\kern-.3em{\rm D}$}}
\def\IE{\relax\hbox{$\inbar\kern-.3em{\rm E}$}}
\def\IF{\relax\hbox{$\inbar\kern-.3em{\rm F}$}}
\def\IG{\relax\hbox{$\inbar\kern-.3em{\rm G}$}}
\def\IGa{\relax\hbox{${\rm I}\kern-.18em\Gamma$}}
\def\IH{\relax{\rm I\kern-.18em H}}
\def\IK{\relax{\rm I\kern-.18em K}}
\def\IL{\relax{\rm I\kern-.18em L}}
\def\IP{\relax{\rm I\kern-.18em P}}
\def\IR{\relax{\rm I\kern-.18em R}}
\def\Z{\relax\ifmmode\mathchoice
{\hbox{\cmss Z\kern-.4em Z}}{\hbox{\cmss Z\kern-.4em Z}} {\lower.9pt\hbox{\cmsss Z\kern-.4em Z}}
{\lower1.2pt\hbox{\cmsss Z\kern-.4em Z}}\else{\cmss Z\kern-.4em Z}\fi}
\def\IZ{Z\!\!\!Z}
\def\II{\relax{\rm I\kern-.18em I}}
\def\one{\relax{\rm 1\kern-.25em I}}

\def\CLL{\relax{\CL\kern-.74em \CL}}


\def\CD {{\cal D}}
\def\CE {{\cal E}}
\def\CF {{\cal F}}
\def\CG {{\cal G}}

\def\CL {{\cal L}}
\def\CM {{\cal M}}
\def\CN {{\cal N}}
\def\CO {{\cal O}}

\def\CR {{\cal R}}

\def\CT {{\cal T}}
\def\CU {{\cal U}}

\def\CW {{\cal W}}
\def\CX {{\cal X}}
\def\CY {{\cal Y}}


\def\p{\partial}

\def\tilde{\widetilde}
\def\hat{\widehat}
\def\bar{\overline}


\def\Tr{{\rm Tr}}

\def\p{\partial}

\def\inbar{\,\vrule height1.5ex width.4pt depth0pt}
\def\r{{\rm Re}}
\def\i{{\rm Im}}

\def\a{\alpha}
\def\b{\beta}
\def\g{\gamma}
\def\d{\delta}
\def\e{\epsilon}

\def\m{\mu}
\def\n{\nu}
\def\la{\lambda}
\def\th{\theta}
\def\s{\sigma}
\def\om{\omega}

\def\bar{\overline}

\def\om{\omega}

\def\det{{\rm det}}

\def\Tr{{\rm Tr}}

\def\IH{{\bf H}}

\chardef\tempcat=\the\catcode`\@
\catcode`\@=11
\def\cyracc{\def\u##1{\if \i##1\accent"24 i
    \else \accent"24 ##1\fi }}
\newfam\cyrfam
\font\tencyr=wncyr10
\def\cyr{\fam\cyrfam\tencyr\cyracc}


\def\boxit#1{\vbox{\hrule\hbox{\vrule\kern8pt
\vbox{\hbox{\kern8pt}\hbox{\vbox{#1}}\hbox{\kern8pt}}
\kern8pt\vrule}\hrule}}
\def\mathboxit#1{\vbox{\hrule\hbox{\vrule\kern8pt\vbox{\kern8pt
\hbox{$\displaystyle #1$}\kern8pt}\kern8pt\vrule}\hrule}}


\lref\Moore{
  G.~W.~Moore,
  ``Arithmetic and attractors,''
  arXiv:hep-th/9807087.
}

\lref\HH{
  J.~B.~Hartle and S.~W.~Hawking,
  ``Wave Function Of The Universe,''
  Phys.\ Rev.\ D {\bf 28}, 2960 (1983).}

\lref\ovv{H.~Ooguri, C.~Vafa and E.~Verlinde, ``Hartle-Hawking wave-function for
flux compactifications: the entropic principle,''
hep-th/0502211.}

\lref\osv{H.~Ooguri, A.~Strominger and C.~Vafa, ``Black hole attractors and the
topological string,'' Phys.\ Rev.\ D
{\bf 70}, 106007 (2004) [arXiv:hep-th/0405146].}

\lref\cano{ P.~Candelas and X.~de la Ossa,
  ``Moduli Space Of Calabi-Yau Manifolds,''
  Nucl.\ Phys.\ B {\bf 355}, 455 (1991).}

\lref\attra{ S.~Ferrara, R.~Kallosh and A.~Strominger,
  ``N=2 extremal black holes,''
  Phys.\ Rev.\ D {\bf 52}, 5412 (1995).
}

\lref\Hgeom{
  N.~J.~Hitchin,
  ``The geometry of three-forms \ in six and seven   dimensions,''
  arXiv:math.dg/0010054.
}

\lref\BCOV{
  M.~Bershadsky, S.~Cecotti, H.~Ooguri and C.~Vafa,
  ``Kodaira-Spencer theory of gravity and exact results for quantum string
  amplitudes,''
  Commun.\ Math.\ Phys.\  {\bf 165}, 311 (1994),
  arXiv:hep-th/9309140.
  }

\lref\GP{
Green, Plesser
}

\lref\TT{
 P.~K.~Tripathy and S.~P.~Trivedi,
  ``Non-supersymmetric attractors in string theory,''
  JHEP {\bf 0603}, 022 (2006),  arXiv:hep-th/0511117.
}

\lref\BFMY{  S.~Bellucci, S.~Ferrara, A.~Marrani and A.~Yeranyan,
   ``Mirror Fermat Calabi-Yau threefolds and Landau-Ginzburg black hole
  attractors,'' arXiv:hep-th/0608091.
}

\lref\GKTT{
  A.~Giryavets, S.~Kachru, P.~K.~Tripathy and S.~P.~Trivedi,
  ``Flux compactifications on Calabi-Yau threefolds,''
  JHEP {\bf 0404}, 003 (2004)
  arXiv:hep-th/0312104.
  }

\lref\COGP{
P.~Candelas, X.~C.~De La Ossa, P.~S.~Green and L.~Parkes,
   ``A pair of Calabi-Yau manifolds as an exactly soluble superconformal
  theory,''
  Nucl.\ Phys.\ B {\bf 359}, 21 (1991).
}

\lref\CAF{
A.~Ceresole, R.~D'Auria and S.~Ferrara,
  ``The Symplectic Structure of N=2 Supergravity and its Central Extension,''
  Nucl.\ Phys.\ Proc.\ Suppl.\  {\bf 46}, 67 (1996)
  [arXiv:hep-th/9509160].
}

\lref\FK{
S.~Ferrara and R.~Kallosh,
  ``Supersymmetry and Attractors,''
  Phys.\ Rev.\ D {\bf 54}, 1514 (1996)
  [arXiv:hep-th/9602136].
}

\lref\Strominger{
  A.~Strominger,
  ``Macroscopic Entropy of $N=2$ Extremal Black Holes,''
  Phys.\ Lett.\ B {\bf 383}, 39 (1996)
  [arXiv:hep-th/9602111].
}

\lref\Moh{
  T.~Mohaupt,
  ``Black hole entropy, special geometry and strings,''
  Fortsch.\ Phys.\  {\bf 49}, 3 (2001)
  [arXiv:hep-th/0007195].
}

\lref\CWMdev{
  G.~Lopes Cardoso, B.~de Wit and T.~Mohaupt,
  ``Deviations from the area law for supersymmetric black holes,''
  Fortsch.\ Phys.\  {\bf 48}, 49 (2000)
  [arXiv:hep-th/9904005].
}

\lref\CWMah{
  G.~L.~Cardoso, B.~de Wit and S.~Mahapatra,
  ``Black hole entropy functions and attractor equations,''
  arXiv:hep-th/0612225.
}

\lref\CWMcor{
  G.~Lopes Cardoso, B.~de Wit and T.~Mohaupt,
  ``Corrections to macroscopic supersymmetric black-hole entropy,''
  Phys.\ Lett.\ B {\bf 451}, 309 (1999)
  [arXiv:hep-th/9812082].
}

\lref\CWMmac{
  G.~Lopes Cardoso, B.~de Wit and T.~Mohaupt,
  ``Macroscopic entropy formulae and non-holomorphic corrections for
  supersymmetric black holes,''
  Nucl.\ Phys.\ B {\bf 567}, 87 (2000)
  [arXiv:hep-th/9906094].
}

\lref\CMWarea{
  G.~Lopes Cardoso, B.~de Wit and T.~Mohaupt,
  ``Area law corrections from state counting and supergravity,''
  Class.\ Quant.\ Grav.\  {\bf 17}, 1007 (2000)
  [arXiv:hep-th/9910179].
}

\lref\FKun{
  S.~Ferrara and R.~Kallosh,
  ``Universality of Supersymmetric Attractors,''
  Phys.\ Rev.\ D {\bf 54}, 1525 (1996)
  [arXiv:hep-th/9603090].
}

\lref\CWMstat{
  G.~Lopes Cardoso, B.~de Wit, J.~Kappeli and T.~Mohaupt,
  ``Stationary BPS solutions in N = 2 supergravity with R**2 interactions,''
  JHEP {\bf 0012}, 019 (2000)
  [arXiv:hep-th/0009234].
}

\lref\Sen{
  A.~Sen,
  ``Black hole entropy function and the attractor mechanism in higher
  derivative gravity,''
  JHEP {\bf 0509}, 038 (2005)
  [arXiv:hep-th/0506177].
}

\lref\TT{
  P.~K.~Tripathy and S.~P.~Trivedi,
  ``Non-supersymmetric attractors in string theory,''
  JHEP {\bf 0603}, 022 (2006)
  [arXiv:hep-th/0511117].
}

\lref\KSS{
  R.~Kallosh, N.~Sivanandam and M.~Soroush,
  ``Exact attractive non-BPS STU black holes,''
  Phys.\ Rev.\  D {\bf 74}, 065008 (2006)
  [arXiv:hep-th/0606263].
}

\lref\FKeight{
  S.~Ferrara and R.~Kallosh,
  ``On N = 8 attractors,''
  Phys.\ Rev.\  D {\bf 73}, 125005 (2006)
  [arXiv:hep-th/0603247].
}

\lref\BFGM{
  S.~Bellucci, S.~Ferrara, M.~Gunaydin and A.~Marrani,
  ``Charge orbits of symmetric special geometries and attractors,''
  Int.\ J.\ Mod.\ Phys.\  A {\bf 21}, 5043 (2006)
  [arXiv:hep-th/0606209].
}

\lref\GIJT{
  K.~Goldstein, N.~Iizuka, R.~P.~Jena and S.~P.~Trivedi,
  ``Non-supersymmetric attractors,''
  Phys.\ Rev.\ D {\bf 72}, 124021 (2005)
  [arXiv:hep-th/0507096].
}

\lref\Senhet{
  A.~Sen,
  ``Entropy function for heterotic black holes,''
  JHEP {\bf 0603}, 008 (2006)
  [arXiv:hep-th/0508042].
}

\lref\Kallosh{
  R.~Kallosh,
  ``New attractors,''
  JHEP {\bf 0512}, 022 (2005)
  [arXiv:hep-th/0510024].
}

\lref\Giryavets{
  A.~Giryavets,
  ``New attractors and area codes,''
  JHEP {\bf 0603}, 020 (2006)
  [arXiv:hep-th/0511215].
}

\lref\AE{
  M.~Alishahiha and H.~Ebrahim,
  ``Non-supersymmetric attractors and entropy function,''
  JHEP {\bf 0603}, 003 (2006)
  [arXiv:hep-th/0601016].
}

\lref\KSS{
  R.~Kallosh, N.~Sivanandam and M.~Soroush,
  ``The non-BPS black hole attractor equation,''
  JHEP {\bf 0603}, 060 (2006)
  [arXiv:hep-th/0602005].
}

\lref\CPTY{
  B.~Chandrasekhar, S.~Parvizi, A.~Tavanfar and H.~Yavartanoo,
  ``Non-supersymmetric attractors in R**2 gravities,''
  JHEP {\bf 0608}, 004 (2006)
  [arXiv:hep-th/0602022].
}

\lref\BFM{
  S.~Bellucci, S.~Ferrara and A.~Marrani,
  ``On some properties of the attractor equations,''
  Phys.\ Lett.\ B {\bf 635}, 172 (2006)
  [arXiv:hep-th/0602161].
}

\lref\SS{
  B.~Sahoo and A.~Sen,
  ``Higher derivative corrections to non-supersymmetric extremal black holes in
  N = 2 supergravity,''
  JHEP {\bf 0609}, 029 (2006)
  [arXiv:hep-th/0603149].
}

\lref\AE{
  M.~Alishahiha and H.~Ebrahim,
  ``New attractor, entropy function and black hole partition function,''
  JHEP {\bf 0611}, 017 (2006)
  [arXiv:hep-th/0605279].
 }

\lref\Exirifard{
  G.~Exirifard,
  ``The world-sheet corrections to dyons in the heterotic theory,''
  arXiv:hep-th/0607094.
}

\lref\SSalp{
  B.~Sahoo and A.~Sen,
  ``alpha' corrections to extremal dyonic black holes in heterotic string
  theory,''
  arXiv:hep-th/0608182.
 }

\lref\AGM{
  D.~Astefanesei, K.~Goldstein and S.~Mahapatra,
  ``Moduli and (un)attractor black hole thermodynamics,''
  arXiv:hep-th/0611140.
 }

\lref\DST{
  A.~Dabholkar, A.~Sen and S.~P.~Trivedi,
  ``Black hole microstates and attractor without supersymmetry,''
  arXiv:hep-th/0611143.
}

\lref\LWKM{
  G.~Lopes Cardoso, B.~de Wit, J.~Kappeli and T.~Mohaupt,
  ``Black hole partition functions and duality,''
  JHEP {\bf 0603}, 074 (2006)
  [arXiv:hep-th/0601108].
}

\lref\AAFT{
  L.~Andrianopoli, R.~D'Auria, S.~Ferrara and M.~Trigiante,
  ``Extremal black holes in supergravity,''
  arXiv:hep-th/0611345.
}

\lref\Pioline{
  B.~Pioline,
  ``Lectures on black holes, topological strings and quantum attractors,''
  Class.\ Quant.\ Grav.\  {\bf 23}, S981 (2006)
  [arXiv:hep-th/0607227].
 }

\lref\KalloshBNB{
  R.~Kallosh,
  ``From BPS to non-BPS black holes canonically,''
  arXiv:hep-th/0603003.
}

\lref\BLWLMS{
  K.~Behrndt, G.~Lopes Cardoso, B.~de Wit, D.~Lust, T.~Mohaupt and W.~A.~Sabra,
  ``Higher-order black-hole solutions in N = 2 supergravity and Calabi-Yau
  string backgrounds,''
  Phys.\ Lett.\ B {\bf 429}, 289 (1998)
  [arXiv:hep-th/9801081].
}

\lref\FGK{
  S.~Ferrara, G.~W.~Gibbons and R.~Kallosh,
  ``Black holes and critical points in moduli space,''
  Nucl.\ Phys.\ B {\bf 500}, 75 (1997)
  [arXiv:hep-th/9702103].
}

\lref\CAF{
  A.~Ceresole, R.~D'Auria and S.~Ferrara,
  ``The Symplectic Structure of N=2 Supergravity and its Central Extension,''
  Nucl.\ Phys.\ Proc.\ Suppl.\  {\bf 46}, 67 (1996)
  [arXiv:hep-th/9509160].
}

\lref\WLP{
  B.~de Wit, P.~G.~Lauwers and A.~Van Proeyen,
  ``Lagrangians Of N=2 Supergravity - Matter Systems,''
  Nucl.\ Phys.\ B {\bf 255}, 569 (1985).
}

\lref\AGNT{
 I.~Antoniadis, E.~Gava, K.~S.~Narain and T.~R.~Taylor,
  ``Topological amplitudes in string theory,''
  Nucl.\ Phys.\  B {\bf 413}, 162 (1994)
  [arXiv:hep-th/9307158].
}

\lref\Verlinde{
  E.~P.~Verlinde,
  ``Attractors and the holomorphic anomaly,''
  arXiv:hep-th/0412139.
}

\lref\Zachos{
  C.~K.~Zachos,
  ``Deformation quantization: Quantum mechanics lives and works in
  phase-space,''
  Int.\ J.\ Mod.\ Phys.\  A {\bf 17}, 297 (2002)
  [arXiv:hep-th/0110114].
}

\lref\OVV{
  H.~Ooguri, C.~Vafa and E.~P.~Verlinde,
  ``Hartle-Hawking wave-function for flux compactifications,''
  Lett.\ Math.\ Phys.\  {\bf 74}, 311 (2005)
  [arXiv:hep-th/0502211].
}

\lref\Witten{
E.~Witten,
``Quantum background independence in string theory,''
arXiv:hep-th/9306122.
}

\lref\DVV{
  R.~Dijkgraaf, E.~P.~Verlinde and M.~Vonk,
  ``On the partition sum of the NS five-brane,''
  arXiv:hep-th/0205281.
}

\lref\GNP{
  M.~Gunaydin, A.~Neitzke and B.~Pioline,
  ``Topological wave functions and heat equations,''
  JHEP {\bf 0612}, 070 (2006)
  [arXiv:hep-th/0607200].
}

\lref\SV{
  A.~Strominger and C.~Vafa,
  ``Microscopic Origin of the Bekenstein-Hawking Entropy,''
  Phys.\ Lett.\  B {\bf 379}, 99 (1996)
  [arXiv:hep-th/9601029].
}

\lref\MSW{
  J.~M.~Maldacena, A.~Strominger and E.~Witten,
  ``Black hole entropy in M-theory,''
  JHEP {\bf 9712}, 002 (1997)
  [arXiv:hep-th/9711053].
}

\lref\Peet{
  A.~W.~Peet,
  ``TASI lectures on black holes in string theory,''
  arXiv:hep-th/0008241.
}

\lref\DMW{
  J.~R.~David, G.~Mandal and S.~R.~Wadia,
  ``Microscopic formulation of black holes in string theory,''
  Phys.\ Rept.\  {\bf 369}, 549 (2002)
  [arXiv:hep-th/0203048].
}

\lref\DM{
  F.~Denef and G.~W.~Moore,
  ``Split states, entropy enigmas, holes and halos,''
  arXiv:hep-th/0702146.
}

\lref\DDMP{
  A.~Dabholkar, F.~Denef, G.~W.~Moore and B.~Pioline,
  ``Exact and asymptotic degeneracies of small black holes,''
  JHEP {\bf 0508}, 021 (2005)
  [arXiv:hep-th/0502157].
}

\lref\DDMPlong{
  A.~Dabholkar, F.~Denef, G.~W.~Moore and B.~Pioline,
  ``Precision counting of small black holes,''
  JHEP {\bf 0510}, 096 (2005)
  [arXiv:hep-th/0507014].
}

\lref\SY{
  D.~Shih and X.~Yin,
  ``Exact black hole degeneracies and the topological string,''
  JHEP {\bf 0604}, 034 (2006)
  [arXiv:hep-th/0508174].
}

\lref\BD{
  B.~Bates and F.~Denef,
  ``Exact solutions for supersymmetric stationary black hole composites,''
  arXiv:hep-th/0304094.
}

\lref\Duff{
  M.~J.~Duff,
  ``String triality, black hole entropy and Cayley's hyperdeterminant,''
  arXiv:hep-th/0601134.
}

\lref\Rotating{
  D.~Astefanesei, K.~Goldstein, R.~P.~Jena, A.~Sen and S.~P.~Trivedi,
  ``Rotating attractors,''
  JHEP {\bf 0610}, 058 (2006)
  [arXiv:hep-th/0606244].
}

\lref\CGLP{
  G.~L.~Cardoso, V.~Grass, D.~Lust and J.~Perz,
  ``Extremal non-BPS black holes and entropy extremization,''
  JHEP {\bf 0609}, 078 (2006)
  [arXiv:hep-th/0607202].
}

\lref\PT{
  S.~Parvizi and A.~Tavanfar,
  ``Partition function of non-supersymmetric black holes in the supergravity
  limit,''
  arXiv:hep-th/0602292.
}

\lref\KSSexact{
  R.~Kallosh, N.~Sivanandam and M.~Soroush,
  ``Exact attractive non-BPS STU black holes,''
  Phys.\ Rev.\  D {\bf 74}, 065008 (2006)
  [arXiv:hep-th/0606263].
 }

\lref\CDKL{
  A.~Castro, J.~L.~Davis, P.~Kraus and F.~Larsen,
  ``5D attractors with higher derivatives,''
  arXiv:hep-th/0702072.
 }

\lref\KL{
  P.~Kraus and F.~Larsen,
  ``Microscopic black hole entropy in theories with higher derivatives,''
  JHEP {\bf 0509}, 034 (2005)
  [arXiv:hep-th/0506176].
}

\lref\Nekrasov{
  N.~A.~Nekrasov,
  ``Seiberg-Witten prepotential from instanton counting,''
  Adv.\ Theor.\ Math.\ Phys.\  {\bf 7}, 831 (2004)
  [arXiv:hep-th/0206161].
}

\lref\NO{
  N.~Nekrasov and A.~Okounkov,
  ``Seiberg-Witten theory and random partitions,''
  arXiv:hep-th/0306238.
}

\lref\LMN{
  A.~S.~Losev, A.~Marshakov and N.~A.~Nekrasov,
  ``Small instantons, little strings and free fermions,''
  arXiv:hep-th/0302191.
}

\lref\Nloc{
N.~A.~Nekrasov,
``Localizing gauge theories,''
preprint IHES/P/03/66 ~~~~~~~~~~~~~~~~~~~~~~ \break
[http://www.ihes.fr/PREPRINTS/P03/P03-66.ps.gz]
}

\lref\NY{
H.~Nakajima, K.~Yoshioka,
``Lectures on Instanton Counting,''
[math.AG/0311058]
}

\lref\twoym{
  C.~Vafa,
  ``Two dimensional Yang-Mills, black holes and topological strings,''
  arXiv:hep-th/0406058.
}

\lref\AOSV{
  M.~Aganagic, H.~Ooguri, N.~Saulina and C.~Vafa,
  ``Black holes, q-deformed 2d Yang-Mills, and non-perturbative topological
  strings,''
  Nucl.\ Phys.\  B {\bf 715}, 304 (2005)
  [arXiv:hep-th/0411280].
 }

\lref\ANV{
  M.~Aganagic, A.~Neitzke and C.~Vafa,
  ``BPS microstates and the open topological string wave function,''
  arXiv:hep-th/0504054.
}

\lref\KKV{
  S.~H.~Katz, A.~Klemm and C.~Vafa,
  ``Geometric engineering of quantum field theories,''
  Nucl.\ Phys.\  B {\bf 497}, 173 (1997)
  [arXiv:hep-th/9609239].
}

\lref\KMV{
  S.~Katz, P.~Mayr and C.~Vafa,
  ``Mirror symmetry and exact solution of 4D N = 2 gauge theories. I,''
  Adv.\ Theor.\ Math.\ Phys.\  {\bf 1}, 53 (1998)
  [arXiv:hep-th/9706110].
}

\lref\LN{
  A.~E.~Lawrence and N.~Nekrasov,
  ``Instanton sums and five-dimensional gauge theories,''
  Nucl.\ Phys.\  B {\bf 513}, 239 (1998)
  [arXiv:hep-th/9706025].
}

\lref\Nfive{
  N.~Nekrasov,
  ``Five dimensional gauge theories and relativistic integrable systems,''
  Nucl.\ Phys.\  B {\bf 531}, 323 (1998)
  [arXiv:hep-th/9609219].
}

\lref\Nlect{
  N.~A.~Nekrasov,
  ``Lectures on nonperturbative aspects of supersymmetric gauge theories,''
  Class.\ Quant.\ Grav.\  {\bf 22}, S77 (2005).
 }

\lref\Nstrings{
N.~A.~Nekrasov,
 ``Master Field of N=2 Theories: Derivation of Seiberg-Witten' Solution,''
talk given at Strings 2003, ~~~~~~~~~~~~~~~~~~~~~~~~~~~~~~~~~~~~~~~~~~~~~~~~~~~\break
http://www.yukawa.kyoto-u.ac.jp/contents/seminar/archive/2003/str2003/talks/nekrasov.pdf
}

\lref\HSV{
  J.~J.~Heckman, J.~Seo and C.~Vafa,
  ``Phase structure of a brane/anti-brane system at large N,''
  arXiv:hep-th/0702077.
}

\lref\TM{
  R.~Dijkgraaf, S.~Gukov, A.~Neitzke and C.~Vafa,
  ``Topological M-theory as unification of form theories of gravity,''
  arXiv:hep-th/0411073.
}

\lref\Nikita{
  N.~Nekrasov,
  ``A la recherche de la m-theorie perdue. Z theory: Chasing m/f theory,''
  arXiv:hep-th/0412021.
}

\lref\AKMV{
  M.~Aganagic, A.~Klemm, M.~Marino and C.~Vafa,
  ``The topological vertex,''
  Commun.\ Math.\ Phys.\  {\bf 254}, 425 (2005)
  [arXiv:hep-th/0305132].
  }

\lref\Iqbal{
  A.~Iqbal,

  ``All genus topological string amplitudes and 5-brane webs as Feynman
  diagrams,''
  arXiv:hep-th/0207114.
}

\lref\IKV{
  A.~Iqbal, C.~Kozcaz and C.~Vafa,
  ``The refined topological vertex,''
  arXiv:hep-th/0701156.
}

\lref\IK{
  A.~Iqbal and A.~K.~Kashani-Poor,
  ``Instanton counting and Chern-Simons theory,''
  Adv.\ Theor.\ Math.\ Phys.\  {\bf 7}, 457 (2004)
  [arXiv:hep-th/0212279].
}

\lref\IKgeom{
  A.~Iqbal and A.~K.~Kashani-Poor,
  ``SU(N) geometries and topological string amplitudes,''
  Adv.\ Theor.\ Math.\ Phys.\  {\bf 10}, 1 (2006)
  [arXiv:hep-th/0306032].
}

\lref\HIV{
  T.~J.~Hollowood, A.~Iqbal and C.~Vafa,
  ``Matrix models, geometric engineering and elliptic genera,''
  arXiv:hep-th/0310272.
}

\lref\EK{
  T.~Eguchi and H.~Kanno,
  ``Topological strings and Nekrasov's formulas,''
  JHEP {\bf 0312}, 006 (2003)
  [arXiv:hep-th/0310235].
}

\lref\Tachikawa{
  Y.~Tachikawa,
  ``Five-dimensional Chern-Simons terms and Nekrasov's instanton counting,''
  JHEP {\bf 0402}, 050 (2004)
  [arXiv:hep-th/0401184].
}

\lref\Wdual{
  E.~Witten,
  ``On S duality in Abelian gauge theory,''
  Selecta Math.\  {\bf 1}, 383 (1995)
  [arXiv:hep-th/9505186].
}

\lref\MW{
  G.~W.~Moore and E.~Witten,
  ``Integration over the u-plane in Donaldson theory,''
  Adv.\ Theor.\ Math.\ Phys.\  {\bf 1}, 298 (1998)
  [arXiv:hep-th/9709193].
}

\lref\EH{
  R.~Emparan and G.~T.~Horowitz,
  ``Microstates of a neutral black hole in M theory,''
  Phys.\ Rev.\ Lett.\  {\bf 97}, 141601 (2006)
  [arXiv:hep-th/0607023].
}

\lref\CT{
  M.~Cvetic and A.~A.~Tseytlin,
  ``Solitonic strings and BPS saturated dyonic black holes,''
  Phys.\ Rev.\  D {\bf 53}, 5619 (1996)
  [Erratum-ibid.\  D {\bf 55}, 3907 (1997)]
  [arXiv:hep-th/9512031].
}

\lref\CU{
  M.~Cvetic and D.~Youm,
  ``Dyonic BPS saturated black holes of heterotic string on a six torus,''
  Phys.\ Rev.\  D {\bf 53}, 584 (1996)
  [arXiv:hep-th/9507090].
}

\lref\KauraMV{
  P.~Kaura and A.~Misra,
  ``On the existence of non-supersymmetric black hole attractors for
  two-parameter Calabi-Yau's and attractor equations,''
  Fortsch.\ Phys.\  {\bf 54}, 1109 (2006)
  [arXiv:hep-th/0607132].
}

\lref\BFFL{
  M.~Billo, M.~Frau, F.~Fucito and A.~Lerda,
  ``Instanton calculus in R-R background and the topological string,''
  JHEP {\bf 0611}, 012 (2006)
  [arXiv:hep-th/0606013].
}


\Title{\vbox{\baselineskip11pt\hbox{hep-th/0703214}
\hbox{HUTP-07/A0003}
\hbox{ITEP-TH-5/07}
}}
{\vbox{
\centerline{Non-supersymmetric Black Holes and Topological Strings}
}}
\centerline{
Kirill Saraikin$^{1}$\footnote{$^{\dagger}$}{On leave from: ITEP,
Moscow, 117259, Russia and L.D.Landau ITP, Moscow, 119334, Russia}
and Cumrun Vafa$^{1,2}$}
\medskip
\medskip
\medskip
\vskip 8pt
\centerline{$^1$ \it Jefferson Physical Laboratory, Harvard University,
Cambridge, MA 02138, USA}
\medskip
\centerline{$^2$ \it Center for Theoretical Physics, MIT, \
Cambridge, MA 02139, USA}
\medskip
\medskip
\medskip
\noindent
We study non-supersymmetric, extremal 4 dimensional black holes which arise
upon compactification of type II superstrings on Calabi-Yau threefolds.
We propose a generalization of the OSV conjecture for higher derivative
corrections to the non-supersymmetric black hole entropy, in terms of the
one parameter refinement of topological string introduced by Nekrasov.
We also study the attractor mechanism for non-supersymmetric black holes
and show how the inverse problem of fixing charges in terms of the
attractor value of CY moduli can be explicitly solved.

\medskip
\Date{March 2007}

\listtoc\writetoc

\newsec{Introduction}

String theory provides microscopic description  of the
entropy of certain types of black holes
through the  counting of D-brane bound states.
The predictions
of string theory include not only a confirmation of the
leading semi-classical entropy formula of Bekenstein and Hawking
which was first confirmed in \SV\ (see, e.g. \refs{\Peet,\DMW}
for a review and references)
but also all the subleading quantum gravitational corrections
which was proposed in \osv\ (building on the work of
\refs{\CWMcor, \CWMdev,\CWMmac, \CMWarea, \CWMstat}).
These higher derivative corrections have been confirmed
by explicit microscopic enumeration in a number of
examples \refs{\twoym,\AOSV,\DDMP,\ANV,\DDMPlong,\SY}.

An important feature of extremal black hole solutions in
$\CN=2,4,8$ supergravity in four space-time dimensions is that
some of the scalar fields (lowest components of the vector multiplets)
acquire fixed values at the horizon. These values are determined
by the magnetic and electric  charges $(p^I,q_I)$ of the black hole and
does not depend on the asymptotic values of the fields at infinity.
The so-called attractor mechanism, which is responsible for such fixed point behavior
of the solutions, was first studied in \refs{\attra,\Strominger,\FK,\FKun}
in the context of the BPS black holes
in the leading semiclassical approximation.
Later, the attractor equations describing these fixed points for
BPS black hole solutions
were generalized to incorporate the
higher derivative corrections to $\CN=2$ supergravity Lagrangian
(see \Moh\ for a review).

Using these supergravity results OSV \osv\ conjectured
a simple relation of the form $Z_{\rm BH} = |Z_{\rm top}|^2$
between the (indexed) entropy of a four-dimensional
BPS black hole in a Type II  string Calabi-Yau compactification,
and topological string partition function, evaluated at the attractor
point on the moduli space. Viewed as an asymptotic expansion
in the limit of large  black hole charges, this relation
predicts all order perturbative contributions to the black hole
entropy due to the $F$-term  corrections
in the effective $\CN=2$ supergravity Lagrangian.
Over the last few years,
this led to a significant progress in understanding
the spectrum of D-brane BPS states on
compact and non-compact Calabi-Yau manifolds, and gave
new insights on the topological strings and  quantum cosmology.
For a recent review and references on this subject, see \Pioline.

Define a mixed  partition function for a black hole
with magnetic charge $p^I$
and electric potential $\phi^I$  by
\eqn\nsc{
Z_{\rm BH}(p^I,\phi^I)=\sum_{q_I}\Omega(p^I,q_I) \, e^{-\phi^I q_I},
}
where $\Omega(p^I,q_I)$ represent
supersymmetric black hole degeneracies
for a given set of charges~$(p^I,q_I)$.
Then
the OSV conjecture \osv\ reads
\eqn\osvcn{
\Omega(p^I,q_I) = \int d\phi^I e^{q_I \phi^I}
\big|Z_{\rm top}(p^I,\phi^I)\big|^2 .
}
As was already mentioned in \osv, expression \osvcn\ needs some
additional refinement.
In particular, rigorous definition of \osvcn\
requires taking care of the
background dependence of the topological string partition
function $Z_{\rm top}$, governed by the holomorphic anomaly~\BCOV.
Also, the integration measure, as well as the choice of a suitable
integration contour needs to be specified.
Some of these issues were
investigated in \refs{\DDMP,\DDMPlong,\SY,\Verlinde,\ovv},
see \DM\ for a recent discussion of these and other subtleties.

In this paper we will address an even more fundamental ambiguity
in \osvcn\ that is present
already at the semiclassical level (without
considering higher genus topological string corrections).
The problem is that although the right hand side of \osvcn\
can be defined for any set of charges $(p^I,q_I)$,
it is well known \Moore\ that
not for all such $(p^I,q_I)$
a supersymmetric spherically symmetric black  hole solution exists.
Typically, there is a real codimension one `discriminant'
hypersurface $\CD(p^I,q_I)=0$ in the space of charges,
such that supersymmetric black hole solutions exist only  when
$\CD(p^I,q_I)<0$.
Therefore, in this case $\Omega(p^I,q_I)$ on the left hand side of \osvcn,
representing
a suitable index of BPS states of charge $(p^I,q_I)$, is zero.

This phenomenon can be illustrated by several examples.
Consider compactification of Type IIB string theory on
the diagonal
$T^6=\Sigma_{\tau}\times\Sigma_{\tau}\times \Sigma_{\tau}$ \refs{\Moore,\BD},
where $\Sigma_{\tau}$ is the elliptic curve with modular parameter $\tau$,
with $D3$-brane wrapping a real 3-cycle on $T^6$.  This can be viewed
as part of the Calabi-Yau moduli when we orbifold $T^6$.  In this paper when
we refer to the diagonal $T^6$ we have in mind the corresponding locus in the
moduli of an associated Calabi-Yau
3-fold with $\CN=2$ supersymmetry where part of the homology of the CY 3-cycles is identified with
the charges $(p^I,q_I)$.
Let the charge configuration be invariant under the
permutation symmetry of the three elliptic curves $\Sigma_{\tau}$.
Note also that the diagonal $T^6$ model is a good approximation to the
generic behavior of Type IIB compactification
on a one-modulus Calabi-Yau threefold in the large
radius limit. If we
label homology of 3-cycles on $T^6$ according to the
 mirror IIA $D$-brane charges as $(u,q,p,v)=(D0,D2,D4,D6)$,
the leading contribution
to the corresponding black hole degeneracy
takes the form
\eqn\omtsix{
\Omega_{\rm susy}(p,q,u,v) \approx
\exp \Big(\pi \sqrt{-\CD(p,q,u,v)}\Big),
}
where the discriminant is
$\CD(p,q,u,v)=-\big(3 p^2 q^2+4 p^3 u+4 q^3 v+6 p q u v-u^2 v^2)$.
It is clear that for some sets of charges this quartic polynomial
can become positive (for example, it is always
the case for $D0-D6$ system, where $\CD(0,0,u,v)=u^2 v^2$),
and \omtsix\ breaks down.
Similar situations occurs in $\CN=2$ truncation of the
heterotic string on $T^6$, the so-called $STU$ model, where $\CD$ becomes
Cayley's hyperdeterminant \Duff\ that can also be
either positive or negative.
Another example of this phenomenon arises in Type IIB compactification  on
$K3\times T^2$. This leads to $\CN=4$ supergravity in four dimensions,
and corresponding expression for the degeneracy \refs{\FKun,\CU,\CT}
\eqn\omktrtt{
\Omega_{\rm susy}(p^I,q_I) \approx
\exp \Big( \pi \sqrt{(P\cdot P) (Q\cdot Q)-(P\cdot Q)^2}\Big).
}
breaks down when $(P\cdot Q)^2 > (P\cdot P) (Q\cdot Q)$.

Thus, the OSV formula \osvcn\ needs to be modified even at the
semiclassical level. One remedy one may think is to sum in \nsc\
only over the charges that support BPS  states: $Z_{\rm
BH}(p^I,\phi^I)= \sum\limits_{q_I: \CD(p^I,q_I)\leq 0}\Omega_{\rm
susy}(p^I,q_I) \, e^{-\phi^I q_I}$. This, however, will not work
because the inverse transform of the topological string partition
function would have to automatically give zero when $(p^I,q_I)$
are non-supersymmetric.  This however turns out not to be the
case, and one gets the naive analytic continuation of the BPS case
(leading to imaginary entropy!). Instead, we can use an
observation that in many examples studied recently in the
literature \refs{\TT,\KSS,\KSSexact,\FKeight,\BFGM,\BFMY} there
exists a non-supersymmetric extremal black hole solution for those
sets of charges that do not support a BPS  black hole:
$\CD(p^I,q_I)>0$. The attractor behavior of a non-supersymmetric
extremal black hole solutions
\refs{\Kallosh,\GIJT,\Giryavets,\BFM,\KalloshBNB,\KauraMV,\CGLP,\AGM,\AAFT}
is similar to the BPS black hole case, since it is a consequence
of extremality rather than supersymmetry~\FGK. Moreover, in the
simplest examples, the macroscopic entropy of a non-supersymmteric
extremal black holes is proportional to the square root of the
discriminant: $S_{\rm BH}^{\rm n-susy}\approx \pi \sqrt{\CD}$, so
that a general expression for the extremal black holes degeneracy
takes the form
\eqn\omextrm{
\Omega_{\rm extrm}(p^I,q_I) \approx
\exp \Big({\pi \sqrt{\big|\CD(p^I,q_I)\big|}}\, \Big),
}
which is valid both for supersymmetric $\CD\leq 0$
and non-supersymmetric $\CD>0$ solutions.

Therefore, it is natural to look for
an extension of the OSV formula \osvcn\ that
can be applied {\it simultaneously} for both BPS and non-BPS extremal
black holes and obtain
corrections to their entropy due to higher derivative terms
in the Lagrangian as a perturbative series in the
inverse charge.
Recently, several steps in this direction were taken from the supergravity
side. A general method (the entropy function formalism)
for computing the macroscopic entropy of
extremal black holes based on $\CN=2$ supergravity action in the presence of
higher-derivative interactions was developed in \refs{\Sen,\Senhet},
and applied for studying corrected attractor equations
and corresponding entropy formula for non-supersymmetric black holes in
\refs{\CPTY,\SS,\AE,\Rotating,\Exirifard,\SSalp,\LWKM,\DST,\CWMah}.
A five-dimensional viewpoint on higher derivative corrections
to attractor equations and entropy, based on
the $c$-function extremization, was developed in \refs{\KL,\CDKL}.
Black hole partition function for non-supersymmetric
extremal black holes was discussed in \refs{\AE,\PT}.

In this paper  we  propose a generalization of \osvcn\ motivated
by the topological string considerations as well as the work \SS :
It was observed in \SS\ that the higher order corrections to the
non-supersymmetric black hole entropy needs higher derivative
corrections in the $\CN=2$ theory which are not purely
antiself-dual in the 4d sense, because unlike the BPS case, the
radii of $AdS_2$ and $S^2$ factors of the near horizon geometry
are not the same. Thus, more information than $F$-terms computed by
topological strings, which only capture antiself-dual geometries,
is needed.  Indeed if one considers only the antiself-dual higher
derivative corrections to the 4d action, there is already a
contradiction with the microscopic count of the non-supersymmetric
black hole at one loop~\SS. Instead it is natural to look for  an
extension of topological string which incorporates
non-antiself-dual corrections as well.  Such a generalization of
topological strings, in the context of geometrically engineered
gauge theories have been proposed by Nekrasov \Nekrasov , where
the string coupling constant is replaced by a pair of parameters
$(\epsilon_1,\epsilon_2)$ which roughly speaking capture the
strength of the graviphoton field strength in the 12 and 34
directions of the 4d non-compact spacetime respectively.  In the
limit when $\epsilon_1=-\epsilon_2=g_{\rm top}$ one recovers back
the ordinary topological string expansion.  However when
$\epsilon_1\not =-\epsilon_2$ this refinement of the topological
string partition function computes additional terms in the 4d
effective theory, as appears to be needed for a correct accounting
of the entropy for non-supersymmetric black holes. This includes a
term proportional to $\CR^2$ which as was found in \SSalp\ is
needed to get the correct one loop correction which is captured by
the refined topological string partition function, but not the
standard one.

Motivated
by this observation and identifying $(\epsilon_1,\epsilon_2)$ with physical fluxes in the
non-supersymmetric black
hole geometry, and motivated by the computations in \SS\ we propose a conjecture for the partition function
of an OSV-like ensemble which includes both BPS and non-supersymmetric
extremal black holes.  We conjecture
\eqn\gosvcnj{
\Omega_{\rm extrm}(p^I,q_I) = \int d\phi^I e^{q_I \phi^I}
\sum_{\rm susy,n-susy} \Big|e^{ {i\pi\over 2} \CG(p^I,\phi^I)}\Big|^2,
}
where   $\CG(p^I,\phi^I)$ is obtained from the $\CG$-function
\eqn\gfdfn{\eqalign{
\CG=
{1\over 2}\big(P^I_{\e}-X^I\big) \big(P^J_{\e}-X^J\big)
\bar F_{IJ}(\bar X,\bar \e) +
\big(P^I_{\e}-X^I\big) F_{I}( X,\e) + F( X,\e)+ ~~~~~~ & \cr
+ {1 \over 2  } (\e_1+\e_2) \bar X{^I}F_{I}( X,\e)-
{1 \over 2} (\e_1+\e_2)\big(\e_1 \p_{\e_1} -\e_2 \p_{\e_2}\big) F( X,\e)+
\CO(\e_1+\e_2)^2&, \cr
P^I_{\e}=-\e_2 p^I +{i\over \pi}\e_1 \phi^I &,
}}
by extremizing $ \i \CG$ with respect to
the parameters $\e_{1,2}$ and (extended) Calabi-Yau moduli~$X^I$,
and  then substituting corresponding solution
$\e_{1,2}=\e_{1,2}(p,\phi),\  X^I=X^I(p,\phi)$ back into $\CG$ \gfdfn .
The sum  in \gosvcnj\ is  over all such solutions to the extremum equations
$\p_{\e_{1,2}}   \i \CG=\p_I  \i \CG=0$,  one of which
ends up being the supersymmetric one given by
$X^I(p,\phi)=p^I+{i\over \pi}\phi^I$, reproducing the OSV conjecture for this case.
The function  $F( X,\e)\equiv F( X^I,\e_1,\e_2)$ in  \gfdfn\ denotes Nekrasov's refinement of the
topological string free energy\foot{
Supersymmetric solution corresponds to $\e_1=-\e_2=1$,
in this case we use the same conventions as in \osv,
and  find $\CG_{\rm susy}(p^I,\phi^I)\equiv F(p^I+{i\over \pi}\phi^I,256)$.
Nekrasov's extension of the
topological string is discussed in section 7.1 below.}.
Depending on the choice of the charges~$(p^I,q_I)$, integration over $\phi^I$ near
the saddle point picks out  supersymmetric or non-supersymmetric black hole
solution.
In the supersymmetric case it reduces to the OSV formula.
In the non-supersymmetric case the corrections have the general structure
suggested by \SS\ (however the exact match cannot be made because
\SS\ only consider higher derivative terms captured by standard topological string
corrections).

The above conjecture is the minimal extension of OSV needed to incorporate
non-supersymmetric corrections.  It is conceivable that there are further
$O(\epsilon_1+\epsilon_2)^2$ corrections to this conjecture.  Such corrections
will not ruin the fact that supersymmetric saddle point still reproduces
the OSV conjecture.

The rest of the paper is organized as follows:
In section 2 we review the attractor equations and
entropy formula for supersymmetric and non-supersymmetric extremal
black holes of $d=4$, $\CN=2$ supergravity arising in the leading
semiclassical approximation. In section 3 we discuss
an alternative formulation of the attractor equations which helps us to
treat supersymmetric and non-supersymmetric black holes in a unified way,
suitable for using in an OSV-like formula.
In section 4 we formulate the {\it inverse problem}  that
allows us to find magnetic and electric
charges of the extremal black hole in terms of the
values of the moduli in vector multiplets fixed at the horizon.
We  give a solution to this problem
for a general one-modulus Calabi-Yau compactification.
In section 5 we discuss semiclassical approximation
to the generalized OSV formula for extremal black holes.
In section 6 we review the results \refs{\SS,\AE,\CWMah} for a corrected
black hole entropy in $\CN=2$ supergravity with higher-derivative
couplings, obtained using the entropy function formalizm.
In section 7 we  observe that matching with the supergravity computations requires
replacing the string coupling constant with two variables  on the topological string
side, and
identify these variables as an
equivariant  parameters  in Nekrasov's extension of the topological
string. This allows us to formulate a generalization of the OSV entropy
formula which is conjectured to be valid asymptotically in the limit of large
charges both for the supersymmetric and non-supersymmetric extremal black holes.
We conclude in section 8 with a discussion of our results and
directions for future research.
Appendix~A contains explicit solutions of the inverse and direct problems
relating the charges and corresponding attractor complex structures
for the diagonal $T^6$ model.

\newsec{The Black Hole Potential and Attractors}

Let us review the attractor equations for extremal
black holes in $d=4$, $\CN=2$ supergravity,
arising in the context of type IIB compactification
on a Calabi-Yau manifold~$M$.
We start by choosing a symplectic basis of 3-cycles
$(A^I, B_I)_{I=0,  \ldots h^{2,1}}$ on $M$,
such that
\eqn\xfdf{
X^I= \int_{A^I} \Omega, \qquad F_I=\p_I F= \int_{B_I} \Omega,
}
where $\Omega$ is a holomorphic 3-form and $F$ is the prepotential of
the Calabi-Yau manifold.
We also choose a basis of  3-forms $(\a_I,\b^I) \in H^{3}(M,\IZ)$ dual to $(A^I, B_I)$.
The K\"{a}hler potential is given by\foot{We use the Einstein convention
and always sum over repeated indices in the paper.}
\eqn\kdf{
K(X,\bar X) = - \log \Big( -i \int_M \Omega \wedge \bar \Omega \Big)=
- \log i \big( \bar X{^I} F_I - X^I \bar F{_I} \big).
}
It defines the K\"{a}hler metric $g_{i\bar j} = \p_i \bar \p_{\bar j} K$.
Let us introduce the superpotential
\eqn\wdf{
\CW = \int_M   \Omega   \wedge H,
}
where
\eqn\hpq{
H=p^I \a_I + q_I \b^I
}
is the RR 3-form, parameterized by a set of (integral)
magnetic and electric charges $(p^I,q_I)$.
The central charge is defined by
\eqn\zdf{
Z = e^{K\over 2 } \CW.
}
Attractor points are  the solutions minimizing the
so-called black hole potential \refs{\FK,\FKun,\FGK,\CAF}
\eqn\vdf{
V_{\rm BH} = |Z|^2 + |DZ|^2.
}
Here $D$ is a fully covariant derivative\foot{On the
objects of K\"{a}hler weight $w$ it acts as $D = \p + w \p K +\Gamma$,
where $\Gamma$ is the Levi-Civita connection of the K\"{a}hler metric.
For example, $DZ = \p Z + {1 \over 2} Z \p K$.},
and $|DZ|^2=g^{i\bar j} D_{i} Z \bar D_{\bar j} \bar Z$.
Notice that for a fixed complex structure on  Calabi-Yau
the central charge \zdf\ is linear in
the charges $(p^I, q_I)$, and therefore the
black hole potential \vdf\ is quadratic in the charges.

We are interested in describing the extremum points of the
potential \vdf. These points correspond to the solutions of the
following equations \FGK
\eqn\extrm{\eqalign{
\partial_i V_{\rm BH} = 2 \bar Z D_i Z +g^{k\bar j}
\big(D_i D_k Z\big) \bar D_{\bar j} \bar Z =0,\cr
\bar \partial_{\bar i} V_{\rm BH} = 2  Z \bar D_{\bar i} \bar Z +
g^{j \bar k} \big(\bar D_{\bar i}  \bar D_{\bar k} \bar Z\big)  D_{ j} Z =0.
}}
There are two types of the solutions, which can be
identified as follows. From the second equation in \extrm\ we find, assuming $Z\not =0$
\eqn\dz{
\bar D_{\bar j} \bar Z =
- {g^{l \bar k} \big(\bar D_{\bar j}  \bar D_{\bar k} \bar Z\big)
\over 2 Z}  D_{l} Z.
}
By substituting this into the first equation in \extrm,
we obtain\foot{Similar expression was derived in \BFM, see eq. (3.5).
In fact, it is straightforward to see that up to a term
which annihilates $D_j Z$ due to \extrm,
the matrix ${\rm M}_i^{\ j} $ is the square ${\rm M} \sim M \bar M$
of the matrix $M_{ij}$ used in~\BFM.
Also, note that the matrix ${\rm M}_i^{\ j} $ can be used to
classify the attractor solutions without assuming $Z \not =0$
(see, e.g. \BFGM\ for explicit examples of the
non-supersymmetric attractor solutions with $Z=0$).
We thank S. Ferrara for this  clarification.
}
\eqn\ft{
 {\rm M}_i^{\ j} D_j Z=0,
}
where
\eqn\mmat{
 {\rm M}_i^{\ j} = 4 |Z|^2 \delta_i^{\ j}- \big(D_i D_k Z\big)  g^{k\bar m}
 \big(\bar D_{\bar m}  \bar D_{\bar n} \bar Z\big)
g^{ \bar n j}
}
Now it is clear that there are two types of solutions to \ft:
\eqn\sns{\eqalign{ {\rm susy:}  \quad  & \det {\rm M}\not = 0, \quad
D_i Z=0    \cr
 {\rm non-susy:} \quad & \det {\rm M} = 0, \quad D_i Z=v_i,
}}
where $v_i$ are the null-vectors: ${\rm M}_i^{\ j} v_j=0$ of
the matrix \mmat.

Solutions to the extremum equations \extrm\ minimize the
black hole potential \vdf, if the Hessian
\eqn\hess{\eqalign{
Hess(V_{\rm BH})=\pmatrix{{\p_i \p_j V_{\rm BH} } &
{ \p_i  \bar \p{_{\bar j}} V_{\rm BH} } \cr
{\bar \p{_{\bar i}} \p_j V_{\rm BH} } &
{\bar \p{_{\bar i}} \bar \p{_{\bar j}} V_{\rm BH}  } },
}}
computed at the extremal point,
is positive definite: $Hess(V_{\rm BH})\big|_{\p V_{\rm BH}=0}>0$.
We will refer to such solutions as attractor points.
According to the classification \sns,
these attractors can be  supersymmetric or non-supersymmetric.
It is easy to show that all  supersymmetric solutions \sns\
minimize the black hole potential.
This is, however, not true in general for the non-supersymmetric solutions,
see e.g. \refs{\TT,\BFMY,\BFM} for some examples.

The black hole potential \vdf\ is related to the
Bekenstein-Hawking entropy of the corresponding
black hole in a simple way. In the classical geometry approximation (at the string
tree level) the entropy is just $\pi$ times the
value of the  potential \vdf\ at the attractor point
\eqn\vbhbh{
S_{\rm BH} = \pi V_{\rm BH}\big|_{\p V_{\rm BH}=0}.
}
After appropriate modification of  the black hole potential
this formula gives corrections to the classical Bekenstein-Hawking  entropy
in the presence of higher derivative terms.
This can be effectively realized using the entropy
function formalism \refs{\Sen,\Senhet}.

\newsec{An Alternative Form of the Attractor Equations}

In this section we discuss an alternative form of the
attractor equations describing extremal
black holes in
$d=4,\CN=2$ supergravity coupled to $n_V$ vector multiplets
in the absence of
higher derivative terms.
We describe two versions of attractor equations,
one involving inhomogeneous and another involving homogeneous coordinates
on Calabi-Yau moduli space.
A natural generalization of these equations
in the presence of higher derivative corrections will be  introduced later
in section 7.

It is convenient to start with the following representation of the black
hole potential~\FGK
\eqn\bhpot{
V_{\rm BH} = -{1\over 2} \big(q_I-\CN_{IJ} p^J\big)
 \Big({1\over \i \CN}\Big)^{IJ}  \big(q^J-\bar \CN{^{JK}}  p^K\big),
}
where
\eqn\ndef{
\CN_{IJ}=\bar F_{IJ}+ 2 i {\i \big(F_{IK} \big) X^K
\i \big(F_{JL}\big) X^L   \over \i \big(F_{MN} \big)X^M X^N},
\qquad F_{IJ} = {\p^2 F \over \p X^I \p X^J}.
}
Notice that $\CN_{IJ}$ is $(n_V+1)\times(n_V+1)$
symmetric complex matrix, and $\i \CN_{IJ}$
is a negative definite matrix, as opposed to $\i F_{IJ}$, which is of signature $(1,n_V)$.
This is clear from the following identity~\CAF
\eqn\imn{
-{1\over 2} \Big({1\over \i \CN}\Big)^{IJ}=e^K  \big(
X^I \bar X{^{J}} +g^{i\bar j} D_i X^J \bar D_{\bar j} \bar X{^{J}} \big).
}
One can use \imn\ and the defining relation \CAF\
\eqn\ndf{
F_I=\CN_{IJ} X^J
}
to bring \bhpot\ into the form \vdf.
Indeed, since
\eqn\rlsn{
(q_I-\CN_{JI}p^J)X^I = q_I X^I - p^I F_I=\CW,
}
the black hole potential \bhpot\ takes
the form
\eqn\vvbh{
V_{\rm BH} = e^K \big(\CW \bar \CW + g^{i\bar j}
D_i \CW \bar D{_{\bar j}} \bar \CW\big),
}
which is equivalent to \vdf.

\subsec{Attractor equations and inhomogeneous variables}

Let us  introduce an auxiliary field $P^I$ that
later will be identified with the
complexified magnetic charge $p^I$, and consider a modified
black hole potential
\eqn\vvv{
V_{\rm BH} = {1\over 2} P^I \i  ( \CN_{IJ}  )\bar P^J -
{i\over 2}  P^I(q_I-\CN_{IJ}  p^J)
+ {i\over 2}\bar P{^I}(q_I-\bar \CN{_{JK}} p^K),
}
where $P^I$ serves
as a Lagrange multiplier.
We want to describe the extrema of $V_{\rm BH}$.
Variation of \vvv\ with respect to $\bar P{^I}$ gives
\eqn\vvar{
P^I=-{i\over \i \CN_{IJ}} \big(q_J-\bar \CN{_{JK}} p^K\big).
}
By plugging this expression form $P^I$ back to  \vvv\ we obtain
the original black hole potential~\bhpot .
It is straightforward to solve  equations \vvar\ in terms
of  the charges:
\eqn\chsol{\mathboxit{\eqalign{
p^I &= \r \big(P^I\big) \cr
q_I &= \r \big(\CN_{IJ} P^J\big)
}}}
Variation of \vvv\ with respect to the Calabi-Yau moduli $\p_i V_{\rm BH}=0$ gives
\eqn\vmod{
P^I  \bar P{^J}  \p_i \i \CN_{IJ}+
i \big(P^I  \p_i \CN_{IJ} - \bar P{^J} \p_i  \bar \CN{_{IJ}}\big) p^J = 0.
}
After using the solution \chsol, we obtain
\eqn\yeqn{
P^I \p_i  \CN_{IJ}  P^J -\bar P{^I}  \p_i  \bar \CN{_{IJ}}  \bar P{^J} =0.
}
This set of the extremum equations can also be written in a compact form as follows
\eqn\yexc{\mathboxit{
\p_i \i ( P^I   \CN_{IJ}  P^J) =0.
}}
For a fixed set of charges $(p^I,q_I)$, solutions to the combined system of
equations \chsol\ and \yexc\ which minimize the modified potential \vvv\
 correspond to the extremal black holes.

Among these, there is always a special  solution
of the form
\eqn\susol{
P^I =C X^I,
}
where $C$ is the complex constant.
Indeed, in this case extremum equations \yexc\ read
\eqn\yexcs{
 C^2 X^I X^J  \p_i  \CN_{IJ}- \bar C{^2}\bar X{^I} \bar X{^J}  \p_i  \bar \CN_{IJ}=0.
}
The second term in \yexcs\ vanishes since
$\bar X{^I} \p_i  \bar \CN_{IJ} =\p_i \big(\bar \CN{_{IJ}} \bar X{^J}\big) =
\p_i \bar F{_I}  =0 $
according to \ndf. The first term in \yexcs\ vanishes because of the
special geometry relation
\eqn\sgxdf{
0=\int_M \Omega \wedge \p_i \Omega = X^I \p_i F_I - F_I \p_i X^I=
X^I X^J  \p_i  \CN_{IJ}.
}
The solution \susol\ describes supersymmetric attractors
\refs{\attra,\Strominger,\FK}, since \chsol\
gives in this case the well-known equations
\eqn\suat{\left\{ \eqalign{
p^I &=  \r \big(  C X^I\big) \cr
q_I &=  \r \big( C F_I\big).
}\right.
}
%

\subsec{Attractor equations and homogeneous variables}

Consider the following potential:
\eqn\newac{V_{\rm BH}= q_I \i P^I+ \i (F_{IJ}) \r \big((P^I -X^I)(P^J-X^J)\big)-
{1\over 2}\i \big(F_{IJ} P^I P^J) .
}
We will keep $P^I$ fixed (in particular, $\r P^I=p^I$) and vary
$X^I$. In order to get rid of the scaling of $X^I$ let us
introduce a new variable $T$ by
\eqn\xnv{ X^I =  \hat X^I T,
}
and integrate out $T$ as follows:
\eqn\tint{ e^{\hat V_{\rm BH}} \approx \int dT e^{V_{\rm BH}}.
}
The potential \newac\ is quadratic in $T$
\eqn\relpr{
V_{\rm BH} = q_I \i P^I + \i (F_{IJ}) \r \big(P^I P^J+\hat X^I \hat
X^J T^2-2 \hat X^I P^J T\big)- {1\over 2}\i \big(F_{IJ} P^I P^J) ,
}
since $F_{IJ}$ has zero weight under the rescaling \xnv. Variation
with respect to $T$ gives:
\eqn\tsol{ T = {\hat X^I \i(F_{IJ}) P^J \over \hat X^I \i (F_{IJ})
\hat X^J }
}
Therefore, the semiclassical approximation to \tint\ gives
\eqn\stx{
\hat V_{\rm BH}= q_I \i P^I + {i\over 4} P^I \CN_{IJ}P^J-
{i\over 4}\bar P{^I} \bar \CN{_{IJ}} \bar P{^J},
}
where
\eqn\nmdf{ \CN_{IJ} =\bar F{_{IJ}} + 2 i{  \i(F_{IK})\hat X^K
\i(F_{JL}) \hat X^L \over \hat X^K \i (F_{KL}) \hat X^L }.
}
The expression \stx\ should be compared to the modified black hole potential \vvv,
which reduces to \stx\ if we use $\r P^I=p^I$.

The choice of the potential \newac\ can be motivated by looking at the
$\CN=2$ supergravity action \WLP. At tree level, the coupling of the
vector fields can be described as
\eqn\sugtre{\eqalign{
8\pi  S_{\rm vec}^{\rm tree}= \int d^4 x \Big(
{i \over 4} F_{IJ} \CF^{-I}_{\m\n} \CF^{-J\m\n} +
{1\over 4} \, & \i(F_{IJ}) \bar X{^{J}}  \CF^{-I}_{\m\n}
T^{-\m\n}-\cr
-{1\over 32} &\i(F_{IJ}) \bar X{^{I}} \bar X{^{J}}
T^{-}_{\m\n} T^{-\m\n}
 + h.c. \Big).
}}
Then $V_{\rm BH} - q_I \i P^I$ in \relpr\ can be interpreted as a zero-mode
reduction of \sugtre, with the following identification:
\eqn\zmidt{\eqalign{
\CF^{-I}_{\m\n} &\to i \bar P{^I} \cr
X^I &\to \hat X^I \cr
T^{-}_{\m\n} & \to 4 i \bar T \cr
\int d^4 x & \to 1 .
}}

Let us now discuss the attractor equations that describe the minima of the
modified black hole potential \newac . We can derive them in
two equivalent ways. First, we can vary \stx\ with respect to the
Calabi-Yau moduli, which gives \yexc.
Or, second, we can vary the potential \newac\ with respect
to the homogeneous coordinates $X^I$ {\it before} we integrate out the overall scale~$T$.
This gives $\p_I V_{\rm BH}=0$ and
we obtain the following attractor equations:
\eqn\sxvar{
-{i\over 2} C_{IJK}
\r \big((P^J -X^J)(P^K-X^K)\big) -
\i (F_{IK})\big(P^K-X^K\big)+{i\over 4} C_{IJK} P^J P^K
=0,
}
where
\eqn\cdef{ C_{IJK} = \partial_I F_{JK} =\partial_I \partial_J
\partial_K F.
}
Using the identity
\eqn\cijkz{
C_{IJK}X^K = 0,
}
which follows from the homogeneity relation $X^I F_I = 2F$, we can write
\sxvar\ as
\eqn\nattc{\mathboxit{
C_{IJK}  \big(\bar P{^J} -\bar X{^J}\big)\big(\bar P{^K}-\bar X{^K}\big) =
4 i \i (F_{IJ})\big(P^J-X^J\big)
}}
It is clear that $X^I = P^I$ is the solution of \sxvar.
If we identify $T\to C, \ X^I \to \hat X^I $,
we obtain $P^I = C \hat X^I$, which is the
supersymmetric solution \susol,\suat. Moreover, if we contract \nattc\
with $X^I$ and use \cijkz, we get
\eqn\natom{
\i (F_{IJ})X^I \big(P^J-X^J\big) =0.
}
In the next section will use this relation  to find all other solutions $P^I(X)$ of
the attractor equations \nattc\ in the one-modulus Calabi-Yau case.

\newsec{The Inverse Problem}

For a given set of charges $(p^I,q_I)$ solutions to the system \extrm\ define
the complex structure on~$M$. However, since these  equations are
highly non-linear, it is hard to write down solutions explicitly
for a general Calabi-Yau manifold. On the other hand, since the
black hole potential \vdf\ is quadratic in  charges\foot{This is
clear from looking at the alternative representation \bhpot\
of the black hole potential.} $(p^I,q_I)$,
we can try to solve the inverse problem: For a given point $t^i$
on the Calabi-Yau moduli space, find corresponding set of the charges
$(p^I,q_I)$ that satisfy \extrm.  Similar techniques were used in
\HSV\ to solve the inverse problem for metastable
non-supersymmetric backgrounds in the context of flux compactifications.

\subsec{Inverse problem and inhomogeneous variables}

Strictly speaking, the physical charges $(p^I,q_I)$ are quantized,
but in semiclassical approximation in the limit of large charges we
can ignore this integrality problem and
treat the charges as continuous coordinates.
Another ambiguity in defining the inverse problem
is related to the fact that all sets of charges $(p^I,q_I)$  connected by an
 $Sp(2n_V+2,\IZ)$ transformations give the same point
on the moduli space, since the black hole potential \vdf\ and
hence the extremum equations \extrm\ are symplectically invariant.
Therefore, we need to choose some canonical symplectic basis in $H^{3}(M,\IZ)$ and
keep it fixed.
However, even including that, the inverse problem is not well-defined,
since the extremization of \vdf\
gives only $2n_V$ real equations \extrm\ for $2n_V+2$ real variables $(p^I,q_I)$.
In order to fix this ambiguity, we suggest to look only at the critical points
where the superpotential \wdf\ takes some particular value:
\eqn\wcnr{
\CW=\om,
}
where $\om$ is a new complex parameter. This can be viewed as
a convenient gauge fixing.
Therefore, we are interested in solving the system of equations
\eqn\iprb{
\p_i V_{\rm BH} =\bar \p_{\bar i} V_{\rm BH} = 0, \qquad W=\om.
}
at some particular point $t^i$ on the  Calabi-Yau moduli space.
Then solution of this inverse problem gives a (multivalued)
map: $(t^i, \om)\to (p^I,q_I)$.

Since $\int_M  \Omega  \wedge H = q_IX^I-p^IF_I$, the equation \wcnr\ can be written as
\eqn\wccn{
X^I\big(q_I-\CN_{IJ}p^J\big)=\om.
}
If we then use \chsol, this gives
$
X^I \i (\CN_{IJ}) \bar P{^J} =i \om.
$
Therefore, the solution of the inverse problem
is given by the following system of equations:
\eqn\nateq{{ \eqalign{
p^I &= \r \big( P^I\big) \qquad \qquad \qquad ~
\p_i \i ( P^I  \CN_{IJ}  P^J) =0 \cr
q_I &= \r \big(\CN_{IJ} P^J\big)  \qquad \qquad \,
X^I \i \big(\CN_{IJ} \big) \bar P{^J} = i \om
}}}
In other words, fixing Calabi-Yau moduli and the gauge \wcnr\
allows one to solve for $P^I$ from the two equations on the
right of \nateq. Then the charges are given by the two equations on the left of \nateq.

Among the solutions to \nateq, there always is a supersymmetric solution \susol,
that can be written as
\eqn\susol{
P^I =2 i e^{K}  \bar\om X^I,
}
where we used
$K = -\log \big(- 2 X \cdot \i \CN \cdot \bar X \big )$
to fix the constant $C$ as
\eqn\crel{
C =2 i \bar \om e^{K}=2 i \big(q_I \bar X{^I} -p^I \bar F{_I}\big)e^K
=2 i  \bar Z e^{K\over 2}.
}

An example of the explicit solution of the inverse problem
in the diagonal $T^6$ model is presented in Appendix A.1

\subsec{Inverse problem and homogeneous variables:  one-modulus Calabi-Yau case}

We can think of the homogeneous variables $X^I$ as parameterizing
extended space $\tilde \CM$ of the complex structures on a Calabi-Yau
threefold $M$. This space can also be viewed as a
total space $\tilde \CM$
of the line bundle $\CL \to \CM$ of the holomorphic 3-forms
$H^{3,0}(M,\IC)$ over the Calabi-Yau moduli space
(to be precise, the Teichm\"{u}ller space)~$\CM$.
Let us comment on the dimension of the space of solutions to the system \nattc.
For a fixed extended Calabi-Yau moduli, this is a set of $n_V+1$ complex
quadratic equations  for
$n_V+1$ complex variables~$P^I$.
Therefore, this system can have at most  $2^{n_V+1}$ solutions.
One of them describes supersymmetric black hole and thus there
are at most   $2^{n_V+1}-1$ non-supersymmetric solutions.

Let us discuss the inverse problem for
a one-modulus Calabi-Yau case, when
\eqn\onmpr{
F = (X^0)^2 f(\tau), \qquad \tau = {X^1\over X^0}.
}
The homogeneity relation gives $F_0 = 2 X^0 f -X^1 f'$, where
$f'\equiv \p_{\tau} f$, and we obtain
the following matrix of second derivatives
\eqn\fmonm{
F_{IJ} = \pmatrix{
2f -2 \tau f'  +  \tau^2 f'' & f'-\tau f'' \cr
f'-\tau f''  & f''
}.
}
an the matrix of third derivatives
\eqn\cmonm{
C_{0IJ} = -\tau C_{1IJ}={1\over X^0}\pmatrix{
-\tau^3 f''' & \tau^2 f''' \cr \tau^2 f''' & -\tau f'''
}
}
To simplify  expressions below, let us introduce the notation
\eqn\yvar{
y^I = P^I-X^I.
}
Then the attractor equations \sxvar\ read
\eqn\attsix{\left\{\eqalign{
 C_{0JK} \bar y{^J}\bar y{^K}
 &= 4 i \i (F_{0J}) y^J \cr
C_{1JK} \bar y{^J}\bar y{^K}
&= 4 i \i (F_{1J}) y^J.
}
\right.
}
Using  the relation \cmonm, we find from \attsix
\eqn\nr{
\i (F_{0I}) y^I = -\tau  \i (F_{1I}) y^I,
}
which is equivalent to \natom. To shorten the notations, let us define
\eqn\xlow{
X_I \equiv X^J\i F_{JI}.
}
For example, $X_0\equiv X^0 \i F_{00}+X^1 \i F_{10}$.
Then we find from \nr
\eqn\yon{
y^1 =-{X_0\over  X_1} \, y^0.
}
If we plug this back into \attsix, we obtain
\eqn\yons{
(\bar y{^0} )^2 =\CY y^0,
}
where
\eqn\Ydef{
\CY = -4iX_1
{\big(X^0\big)^4 \det \|\i F_{IJ} \|
\over f''' \big(X^I  X_I \big)^2}
}
For future reference, let us write down an explicit expression for
the ingredients entering~\Ydef, in terms of the holomorphic function $f$
defining the prepotential \onmpr:
\eqn\usfms{\eqalign{
X_1 &=  X^0(\i f'-\i(\tau) \bar{f''}) \cr
X^I X_I  &=2 (X^0)^2 \big(
 \i f -\i(\tau) \bar{f'}-i (\i \tau)^2 \bar{f''}\big) \cr
\det \|\i F_{IJ} \| & = 2 \i(f) \i(f'') -(\i f')^2+
2 \i (\tau) \i (f'\bar {f''})-(\i \tau)^2 |f''|^2.
}}
In order to solve \yons, we take the square of the complex conjugate equation
and then use  \yons. This gives
\eqn\yeqeql{
(y^0)^4 = \bar  \CY^2 \CY y^0.
}
Therefore, in terms of the original
variables \yvar\  we  find the following four solutions:
\eqn\yeqsols{\left\{\eqalign{
P^0_{(0)} &=X^0 \cr
P^1_{(0)} &=X^1,
}\right.
}
and
\eqn\yeqsolns{\left\{\eqalign{
P^0_{(k)} &= X^0+ \big(\bar  \CY^2 \CY\big)^{1/3}e^{2\pi i k/3} \cr
P^1_{(k)} &= X^1-{X_0\over X_1 }
\big(\bar  \CY^2 \CY\big)^{1/3}e^{2\pi i k/3},   \qquad k=1,2,3.
}\right.
}
where the first solution corresponds to a supersymmetric  black hole and the other three are
non-supersymmetric.
Corresponding black hole charges are given by
\eqn\chinp{\left\{\eqalign{
p^I & =  \r P^I \cr
q_I & = \r\big(\CN_{IJ} P^J \big).
}\right.
}
%

\newsec{Semiclassical Entropy in the  OSV Ensamble}

In this section we develop a semiclassical version of OSV formalism which applies
to both supersymmetric  and non-supersymmetric black holes.
We then illustrate it using  $D0-D4$ system in the diagonal $T^6$ model as an example.
This will serve as a preparation for the discussions
in section 6 and the conjecture in section 7 taking into account
perturbative corrections to the extremal black hole entropy.

We begin by recalling some basic ingredients of the OSV formalism.
The formula~\osv
\eqn\osvz{
Z_{\rm BH}(p^I,\phi^I) = \Big|e^{F_{\rm top}(p^I+{i\over \pi}\phi^I )} \Big|^2.
}
describes a relation between the mixed partition function
of the supersymmetric (BPS) black hole  and topological string free energy.
Here $F_{\rm top}$ denotes the topological string free energy.
It is well known~\BCOV\ that
the higher genus contributions to $F_{\rm top}$  depend non-holomorphically on
the background complex structure. This dependence, originally
described in \BCOV\ as the holomorphic anomaly in the topological string
amplitudes coming from the boundary of the moduli space, was interpreted
in~\Witten\ as a dependence of the wave-function $\Psi_{\rm top}= e^{F_{\rm top}}$
on the choice of the polarization.
This viewpoint on the topological string partition function
as a wave-function was further studied in
\refs{\Verlinde, \DVV, \OVV}.

As noted in \osv, the formula \nsc\ can be inverted, and resulting expression
\eqn\osvwig{
\Omega(p^I,q_I) = \int d \chi^I e^{-i\pi \chi^I q_I}
\Psi_{\rm top}^* \big(p^I-\chi^I\big) \Psi_{\rm top}\big(p^I+\chi^I\big).
}
can be interpreted as the Wigner
function\foot{Let us recall that in quantum mechanics the Wigner
function defines the quasi-probablity measure
$f({\rm x},{\rm p}) = {1\over 2\pi} \int dy e^{-iy{\rm p} }
\psi^*({\rm  x}-{\hbar \over 2} y)\psi({\rm  x}+{\hbar \over 2} y)$
on the
{\it phase} space, see e.g. \Zachos. Here the canonical commutation relation
is $[\hat{\rm  p}, \hat{\rm x}]=-i\hbar$. In the topological string setup
$\hbar = {2\over \pi}$.}
associated to
the topological string wave function.
Here $\Psi_{\rm top}(p^I)=\langle p^I |\Psi_{\rm top} \rangle$ represents the
topological string wave function
in real polarization (see \GNP\ for a comprehensive review
and  references),  and the chemical potentials are
restored after deforming the integration contour as $ \phi^I=-i\chi^I $.

\subsec{Black hole potential and OSV transformation}

Let us rewrite modified black hole potential \newac\ in the form
\eqn\newacr{\eqalign{
V_{\rm BH}^{(0)}=q_I \i P^I +\Big({i\over 4} (P^I-X^I) (P^J-X^J) \bar F_{IJ}^{(0)}
+{i\over 2} (P^I-X^I) F_{I}^{(0)}+{i\over 2} F^{(0)}+ c.c.\Big).
}}
We put the superscript $(0)$ to stress that the prepotential $F^{(0)}$ corresponds
to a genus zero part of the topological string free energy.
As in the OSV setup \osv, we can
parameterize the Lagrange multiplier $P^I$ (which can  also be viewed as
a complexified magnetic  charge)~as
\eqn\newacc{
P^I = p^I +{i\over \pi }\phi^I,
}
so that the first of the attractor equations \chsol\ is automatically satisfied.
At the next step, we rewrite the semiclassical entropy
$
S_{\rm BH}^{(0)} =\pi V_{\rm BH}^{(0)}
$
as
\eqn\ssemvbh{
S_{\rm BH}^{(0)}  =q_I  \phi^I - \pi \i \CG^{(0)},
}
where we  introduced a function $\CG^{(0)}$ defined by
\eqn\gfsem{
\CG^{(0)} =
{1\over 2} (P^I-X^I) (P^J-X^J) \bar F_{IJ}^{(0)}
+ (P^I-X^I) F_{I}^{(0)}+ F^{(0)}.
}
In order to compute the entropy in \ssemvbh\  we should find the values of $\phi^I$ and $X^I$
that extremize the black hole potential \newacr .
Extremization  with respect to the (extended) Calabi-Yau moduli
$\p_I V_{\rm BH}^{(0)}=0$ gives the equations \nattc.
Let us use the index $s$ to label all solutions
to these equations, $X^I_s =X^I_s(P)$.
There are two types of these solutions, supersymmetric ($s= \rm susy$)
and non-supersymmetric  ($s=\rm n-susy$) ones.
In particular,  the  supersymmetric
solution is given by $X^I_{\rm susy}(P) =P^I$.
By substituting these solutions in \gfsem\ we obtain the functions
$\CG^{(0)}_s(P^I) = \CG^{(0)}_s(p^I,\phi^I)$.
In the supersymmetric case $\CG^{(0)}_{\rm susy}(P^I) = F^{(0)} (p^I+{i\over \pi}\phi^I)$.
Let us define a mixed partition functions corresponding
to each of the solutions  $X^I_s =X^I_s(P)$   by
\eqn\mxpf{
Z_{s}^{(0)} (p^I,\phi^I) = e^{i {\pi\over 2}\CG^{(0)}_s(p^I,\phi^I)} .
}
For example, the supersymmetric mixed partition function
\eqn\mxpfosv{
Z_{\rm susy}^{(0)} (p^I,\phi^I) =
e^{i {\pi\over 2}F^{(0)} (p^I+{i\over \pi}\phi^I)}
}
describes the leading contribution to \osvz.

For a fixed charge vector $(p^I,q_I)$
the  extremal black hole degeneracy can be written symbolically  as
$\Omega_{\rm extrm} = \Omega_{\rm susy}+\Omega_{\rm n-susy}$.
Therefore, the  leading semiclassical contribution to $\Omega_{\rm extrm}$ is
given by an OSV type integral
\eqn\osvwig{
\Omega^{(0)}_{\rm extrm}(p^I,q^I) =
\int d \phi^I e^{q_I \phi^I} \sum_s \big| Z^{(0)}_{s}(p^I,\phi^I)\big|^2,
}
where the sum is over all  solutions
to the extremum equations \nattc .
We will discuss perturbative corrections to this formula later in section 7,
but before that
let us comment on the possible wave function interpretation of this expression.

Define
\eqn\wfy{
\Psi(X,P) = \exp { i\pi \over 2} \Big({1\over 2} (P^I-X^I) (P^J-X^J) \bar F_{IJ}^{(0)}
+ (P^I-X^I) F_{I}^{(0)}+ F^{(0)}\Big).
}
This is essentially the off-shell version of the partition function \mxpf,
since we have not substituted the extremum solution $X^I_s =X^I_s(P)$ into
\wfy\ yet. This can be achieved by integrating out the fields $X^I$
in the semiclassical approximation
\eqn\pfsemi{
\sum_s \big| Z^{(0)}_{s}(p^I,\phi^I)\big|^2 \approx
\int dX^I d\bar X{^I} \sqrt{\det \|\i F_{IJ} \|} \Psi(X,P)\Psi^*(X,P).
}
The function $\Psi(X,P)$ given in \wfy\  is holomorphic in $P^I$ and non-holomorphic in $X^I$.
It turns out that (up to some numerical factors due to a difference in conventions)
it coincides exactly with the DVV `conformal block'  \DVV\
appearing in study of the five-brane partition function!
In particular, as was shown in \DVV, it satisfies the holomorphic anomaly equation~\BCOV.
Using results of  \GNP, it can be identified  as the intertwining function
$\Psi(X,P) =_{(X, \bar X)}\!\!\langle X^I|P^I\rangle$
between the coherent state $|P^I\rangle$ in the real polarization
and the coherent state $|X^I\rangle_{(X, \bar X)}$
in the holomorphic polarization appearing in quantization of $H^3(M,\IC)$.
The integral in \pfsemi\ then can naturally be interpreted as
averaging over the wave function polarizations, thus effectively removing the background dependence.
We should stress, however, that only semiclassical
approximation to this integral is needed for~\osvwig.
This would be interesting to develop further, especially
in connection with the topological M-theory \refs{\TM,\Nikita}
interpretation of the black hole entropy counting.

We now turn to a simple example of the diagonal $T^6$ model, where
semiclassical formula \osvwig\ for  extremal black hole entropy can be
illustrated.

\subsec{Semiclassical entropy in the diagonal $T^6$ compactification}

Consider Type IIB compactification on the diagonal $T^6$ threefold \Moore\
(see Appendix~A for more details about this model).
The prepotential is
\eqn\tsprp{
F = {(X^1)^3\over X^0}, \qquad f(\tau)=\tau^3,
}
where the complex structure parameter $\tau = {X^1\over X^0}$.
We compute:
\eqn\ftor{
F_{IJ} = \pmatrix{2 \tau^3  & -3\tau^2 \cr
-3 \tau^2 & 6 \tau
}, \quad
C_{IJ0} = -{6 \tau \over X^0}\pmatrix{ \tau^2 & -\tau \cr
- \tau & 1}, \quad
 C_{IJ1}={6  \over X^0}\pmatrix{ \tau^2 & -\tau \cr - \tau & 1}.
}
Let us denote $ y^I = P^I -X^I$.
The attractor equations \nattc\ read
\eqn\attsix{\left\{\eqalign{
 C_{0IJ} \bar y{^I} \bar y{^J}  =& 4 i \i (F_{0I})y^I \cr
C_{1IJ} \bar y{^I} \bar y{^J}  = &4 i \i (F_{1I})y^I.
}
\right.
}
In order to compute the function $\CG^{(0)}(p^I,\phi^I)$, we need
to find from these equations a solution $X^I=X^I(P)$ of the direct problem.
This can be  done by inverting the solutions of the inverse problem \yeqsols-\yeqsolns.
However, it turns out that it is easier to find  $X^I=X^I(P)$ directly from
\attsix.

According to \cmonm\ and \ftor, the third derivatives
of the prepotential are related as $C_{0IJ}=-\tau C_{1IJ}$, and
therefore \attsix\ reduces to
\eqn\atsn{
2y^0 \i (\tau^3)-3 y^1 \i (\tau^2)=3 \tau y^0 \i (\tau^2)-6 \tau y^1 \i (\tau).
}
Apart from the supersymmetric solution $y^0=y^1=0$, this gives
\eqn\nssol{
{y^1\over y^0} =  \r \tau-{i\over 3 }\i \tau,
}
If we recall that $ y^I = P^I -X^I$, we can solve \nssol\ for $X^1$:
\eqn\xonesl{\eqalign{
 X^1=  X^0 {4\r( X^0 \bar P{^1}) -2 \bar P{^1} P^0
 +P^1\bar P{^0} \over 4 \r(X^0 \bar P{^0})-|P^0 |^2} .
}}
Then we plug this into the second equation of \attsix\ and
find\foot{assuming $\i\big(P^0 \bar P{^{1}}\big)\not =0$.}
\eqn\xtwesl{
\big(\bar X{^0}-\bar P{^0}\big)^2=3 X{^0} \big( X{^0} -P^0\big).
}
This should be compared to \yons .
To solve the equation \xtwesl, is convenient to work with
the real and imaginary parts of $X{^0}$ and $P{^0}$.
Then \xtwesl\ can be reduced to a quartic equation for  $\r X{^0}$.
For a generic choice of $\r P{^0}$ and $ \i P{^0}$, two of the roots of this
quartic equation are complex, and
two are real.
These real roots lead to
the two solutions of the attractor equations \attsix, supersymmetric
\eqn\xomsltt{\eqalign{
 \qquad X^0 = P^0, \cr X^1 = P^1,
}}
and non-supersymmetric one. Explicit expression for the non-supersymmetric
solution depends on the signs of $\r P{^0}$ and $ \i P{^0}$. For example,
when $\i P{^0}> |\r P{^0}|$, it is given by\foot{Corresponding solution  for $X^1$ is obtained
by plugging this expression into \xonesl.}
\eqn\xtwsltt{\eqalign{
 \r X^0 = {1\over 4} \r P^0+&
{3\over 8 }\big(\r P^0+ \i P^0\big)^{2\over3} \big(\i P^0- \r P^0\big)^{1\over3}-\cr -&
{3\over 8 } \big(\r P^0+ \i P^0\big)^{1\over3} \big(\i P^0- \r P^0\big)^{2\over3}, \cr
 \i X^0 ={1\over 4} \i P^0-&
{1\over 4}
\sqrt{9 \big(\i P^0\big)^2 -8 \big(\r P^0\big)^2-8 \r\big(X^0\big)\r\big(P^0\big)+16 \big(\r X^0\big)^2}.
}}

We can use these solutions and study
a system of $k D0$ and $N D4$ branes on the diagonal $T^6$.
This corresponds to the charge vector of the form $(k,0,N,0)$.
In this case the discriminant
$
\CD = -(3 p^2 q^2+4 p^3 u+4 q^3 v+6 p q u v-u^2 v^2)
$
reduces to $\CD=-4 k N^3$, so that the system is supersymmetric
when $kN>0$ and non-supersymmetric when $kN<0$.
Complexified magnetic charges are given by
\eqn\ydfdo{
P^0={i\over \pi}\varphi, \qquad  P^1 = N + {i\over \pi} \phi,
}
and the black hole degeneracy \osvwig\ in this case reads
\eqn\omkn{
\Omega_{\rm extrm}^{(0)}(k,N) = \int d \phi d\varphi e^{k\varphi}
\Big(e^{-\pi \i \CG_{\rm susy}^{(0)}({i\over\pi}\varphi,N+{i\over\pi}\phi)}+
e^{-\pi \i \CG_{\rm n-susy}^{(0)} ({i\over\pi}\varphi,N+{i\over\pi}\phi)}\Big).
}

Let us now compute expressions for $\CG^{(0)}$-functions entering into \omkn.
Using \xomsltt, we find from~\gfsem\
\eqn\cgsns{\eqalign{
-\pi \i \CG_{\rm susy}^{(0)} \big({i\over\pi}\varphi,N+{i\over\pi}\phi\big)=
{  N^3 \pi^2- 3 N \phi^2 \over \varphi}.
}}
The non-supersymmetric solution \xtwsltt\ in the case \ydfdo\ reads
\eqn\nnnns{\eqalign{
X^0 = &-{i\over 2\pi}\varphi \cr
X^1=& {1\over 2}\big(N -{i\over 2\pi} \phi \big).
}}
Therefore, from  \gfsem\ we obtain the following expression
\eqn\cgnns{\eqalign{
-\pi \i \CG_{\rm n-susy}\big({i\over\pi}\varphi,N+{i\over\pi}\phi\big)
=-{  N^3 \pi^2- 3 N \phi^2 \over \varphi}.
}}

The integral over $\phi$ in \omkn\ is quadratic, and
(ignoring the convergence issue) in the semiclassical
approximation $\phi=0$ .
The critical points in the $\varphi$ direction are given by
\eqn\phsos{
\p_{\varphi}(k\varphi-\pi \i\CG_{\rm susy})=0 \quad \Rightarrow \quad
\varphi_{\rm susy} = \pi \sqrt{N^3\over k}
}
for supersymmetric term, and
\eqn\phnsos{
\p_{\varphi}(k\varphi-\pi \i\CG_{\rm n-susy})=0 \quad \Rightarrow \quad
\varphi_{\rm n-susy} = \pi \sqrt{-{N^3\over k}}
}
for the non-supersymmetric term. Since we are integrating over the
real axis, the leading contribution to \omkn\
comes only from one of the two terms,  depending on the sign of the
ratio ${N\over k}$. This gives:
\eqn\omknfn{
\Omega_{\rm extrm}^{(0)}(k,N) \approx\exp\big(2 \pi \sqrt{|N^3 k|}\big),
}
which is a correct expression for extremal black hole degeneracy,
valid both in the supersymmetric and non-supersymmetric cases.
Using the same method, it is also easy to obtain an expression
$\Omega_{\rm extrm}^{(0)}(N_0,N_6) \approx\exp\big( \pi |N_0N_6|\big)$
 for the degeneracy of $D0-D6$ system on diagonal $T^6$, which
agrees with \EH.

It is instructive to compare this prediction of \osvwig\ with
the original OSV formula \osv
\eqn\osvrel{
\Omega(p^I,q_I) = \int d\phi^I e^{q_I \phi^I +\CF(p^I,\phi^I)}.
}
Because of our choice of the non-canonical
$D3$-brane intersection matrix (see Appendix A) on $T^6$,
we have
$q_I \phi^I = -u \phi^0 - 3q\phi$. Also,
\eqn\nhfts{
\CF(p^I,\phi^I) = -\pi \i\left({(p+{i\over \pi} \phi)^3\over
v+{i\over \pi} \phi^0}\right).
}
In the semiclassical approximation, the leading contribution to
$\ln \Omega(u,q,p,v)$ can be computed by extremizing the exponent in \osvrel.
This gives
\eqn\exmax{\eqalign{
2q&=-{(p+{i\over \pi} \phi)^2\over v+{i\over \pi} \phi^0} -
{(p-{i\over \pi} \phi)^2\over v- {i\over \pi} \phi^0}, \cr
2u &={(p+{i\over \pi} \phi)^3\over (v+{i\over \pi} \phi^0)^2} -
{(p-{i\over \pi} \phi)^3\over (v-{i\over \pi} \phi^0)^2}.
}}
which  essentially are the supersymmetric attractor equations \suat.
The general solution to \exmax\ is easy to write:
\eqn\exmxsl{\eqalign{
\phi^0&=\pm \pi {2p^3+2pqv -uv^2\over \sqrt{-\CD}}, \cr
\phi &= \mp \pi {2p^2q+2q^2v +puv\over \sqrt{-\CD}},
}}
where the discriminant $\CD=-\big(3 p^2 q^2+4 p^3 u+4 q^3 v+6 p q u v-u^2 v^2)$.
The sign ambiguity in \exmxsl\ can be fixed by imposing physically natural
condition
\eqn\sgnfx{
\i \tau = \i {p+{i\over \pi} \phi\over v+{i\over \pi}\phi^0} >0.
}
Notice that the potentials \exmxsl\ become pure imaginary when $\CD >0$.
Therefore, if one is allowed to do the analytical continuation when
computing the integral \osvrel,  the answer for the microcanonical
entropy reads
\eqn\entin{
\ln \Omega(u,q,p,v)  \approx \pi \sqrt{3 p^2 q^2+4 p^3 u+4 q^3 v+6 p q u v-u^2 v^2}.
}
This expression, of course, becomes pure imaginary on the non-supersymmetric
side $\CD >0$ of the discriminant hypersurface $\CD =0$, which is meaningless.
This thus illustrates the shortcoming of OSV formalism in the context of non-BPS
black holes.

\newsec{Including Higher Derivative Corrections: The Entropy Function Approach}

The Wald's formula provides a convenient tool for computing  the macroscopic
black hole entropy in the presence of
higher derivative terms.
It can be written as
\eqn\swald{
S_{\rm BH} =2 \pi \int_{H} d^2 x \sqrt{h} \,
\e_{\m\n} \e_{\la\rho} {\d \CL \over \d \CR_{\m\n\la\rho}} ,
}
where~$\CL$ is the Lagrangian density and
 the integral is computed over the  black hole horizon.
Sen \refs{\Sen,\Senhet} showed that
in the case of a spherically symmetric extremal black holes with
$AdS^2\times S_2$ near horizon geometry Wald's formula
simplifies drastically.
This gives an effective  method for computing a macroscopic entropy of a
spherically symmetric extremal black holes in a theory of gravity
coupled to gauge and scalar fields, called the
entropy function formalism.

In this section we briefly describe, following~\Moh,
a formulation of
$\CN=2$ supergravity coupled to $n_V$ abelian gauge fields, in the
presence of higher-derivative corrections.
Then we review
recent computations of the
extremal black hole entropy
in this setup  \refs{\SS, \AE, \CWMah},
performed in the framework of the entropy function formalism.

\subsec{ $d=4,\ \CN=2$  Supergravity  with $F$-term $\CR^2$ corrections}

The Lagrangian density of $\CN=2$ Poincare supergravity
coupled to $n_V$ vector multiplets can be
conveniently formulated using the off-shell description \WLP.
The idea is to start with an $\CN=2$ conformal supergravity
and then reduce it to Poincare supergravity by gauge fixing
and adding appropriate compensating fields.
The advantage of working with $\CN=2$  superconformal approach is that it
provides many powerful tools, such as superconformal tensor calculus and
a general {\it density formula} for the Lagrangian.

One introduces the Weil
and matter chiral superfields
\eqn\wmat{\eqalign{
W_{\m\n}(x,\th) &= T_{\m\n}^- -
{1\over 2}\CR_{\m\n\la\rho}^- \e_{\a\b}\th^{\a}\sigma^{\la\rho}\th^{\b}+ \ldots \cr
\Phi^I(x,\th)&=X^I+
{1\over 2}\CF^{-I}_{\m\n} \e_{\a\b}\th^{\a}\sigma^{\m\n}\th^{\b}+\ldots
}
}
where $T^{-}_{\m\n}$ is an auxiliary antiself-dual tensor field\foot{
At tree-level this field is identified with the graviphoton by  the
equations of motion.}, and
$\CF^{-I}_{\m\n}$ and $\CR^{-}_{\m\n\la\rho}$ denote the anti-selfdual
parts  the  field-strength and curvature tensors correspondingly.
The conventions are $* T_{\m\n} = {1\over 2 }\e_{\m\n\rho\s} T^{\rho\s}$
and $T^{\pm}_{\m\n} =  {1\over 2 }(T_{\m\n} \pm i * T_{\m\n})$,
so that $T^{-}_{\m\n} = \bar T{^{+}_{\m\n}}$ for
Minkovski signature.
The superconformally covariant field strength
\eqn\scfinfs{
{\bf F}^{I}_{\m\n} = \CF^{I}_{\m\n}-\big({1\over 4} \bar X{^I} T_{\m\n}^{-}+
\e_{ij}\bar \psi{^i}_{[\m}\g_{\n]}\Omega^{jI}+
\e_{ij} \bar X{^I} \bar \psi{^i}_{\m}\psi^j_{\n}+
h.c.
\big)
}
enters into the bosonic part of the Lagrangian
through the combination $ \CF^{+I}_{\m\n}-{1\over 4} X^I T^{+}_{\m\n}$.
The $F$-terms
can be reproduced from the generalized prepotential
\eqn\genf{
F(X^I,W) =\sum_g F^{(g)}(X^I) W^{2g},
}
where $F^{(g)}$ can be computed
from the topological string amplitudes  \refs{\BCOV,\AGNT}.
In particular, the topological string  free energy is given by
\eqn\ftopdf{
F_{\rm top}(X^I,g_{\rm top}) =\sum_g  (g_{\rm top})^{2g-2}F^{(g)}(X^I).
}
The function $F^{(g)}$ is homogeneous of degree $2-2g$, so that
\eqn\fghomrl{
F(\la X^I,\la W)=\la^2 F(X^I,W).
}
This homogeneity relation for  the
generalized prepotential~\genf\ can also be written as
\eqn\hrgp{
X^I \p_I F + W\p_W F = 2 F.
}
Notice that  another notation
\eqn\fganot{
\hat A \equiv W^2,\qquad F(X^I,\hat A) \equiv F(X^I,W)
}
is sometimes used in the supergravity literature.

The coupling of the vector fields to the gravity is
governed  by the generalized prepotential \genf\ as follows
\eqn\lvct{\eqalign{
8\pi S_{\rm vect}& =8\pi S_{\rm vect}^{\rm tree}+
 \int d^4 x d^4 \th \sum_{g=1}^{\infty}  F_g(\Phi^I) \big(W_{\m\n}W^{\m\n}\big)^g
 + h.c. =\cr
 &= 8\pi S_{\rm vect}^{\rm tree}+  \int d^4 x \sum_{g=1}^{\infty}
 F_g(X^I)\big( \CR^2_{-}T^{2g-2}_{-}
 +\ldots \big) + h.c.
}}
The terms in the Lagrangian density, relevant for the computation of the
entropy are \Moh\
\eqn\slaghc{\eqalign{
8\pi  \CL \! = \!
-{i\over 2}  \Big[
{1\over 2} \big( \CF^{+I}_{\m\n}\! - \! {1\over 4} X^{ I} T^{+}_{\m\n} \big) \!
 \big( \CF^{+J \m\n}\! - \! {1\over 4} X^{ J} T^{+\m\n} \big)\! \bar F{_{IJ}}
\! + \!  { T^{+\m\n} \! \! \over 4} \big( \CF^{+I}_{\m\n}\! -
\! {1\over 4} X^I T^{+}_{\m\n} \big) \! F{_{I}}
+{\hat{\bar{A}}\over 16} F - \! \cr
-\bar X{^I} F_I \CR
- F_{\hat A} \hat C
- h.c. \Big]+\ldots
}}
Here
\eqn\cdef{\eqalign{
\hat C = &64 \CR^{-}_{\n\m\rho\s}\CR^{-\n\m\rho\s} +
16 T^{-\m\n} f^{\rho}_{\m} T^{+}_{\rho\n}+ \ldots \cr
f^{\n}_{\m} = &-{1\over 2} \CR^{\n}_{\m}+
{1\over 32}T^{-}_{\m\rho}T^{+\n\rho} +\ldots \cr
F= & F(X^I, \hat A), \quad F_{\hat A}\equiv \p_{\hat A} F,
}}
and $\ldots$ in \slaghc-\cdef\ denotes the terms (auxiliary fields,
fermions, etc.) that will vanish or cancel out on the black hole ansatz.

\subsec{Review of the entropy function computation}

We are interested in a spherically symmetric
extremal black hole solutions arising in the
supergravity theory defined by the Lagrangian \slaghc.
Consider the most general $SO(2,1)\times SO(3)$ ansatz \SS\
for a field configurations consistent with the
$AdS_2 \times S^2$  near horizon geometry of the black hole
\eqn\senanz{\eqalign{ ds^2 &= v_1\Big( -r^2 dt^2 + {dr^2 \over r^2}
\Big)+ v_2 (d\th^2 + \sin^2 \th d\varphi^2), \cr
X^I &= x^I,\quad \CF^I_{rt} =-{\phi^I\over \pi}, \quad
\CF^I_{\th \varphi} = {p^I  } \sin \th,
\quad T^{-}_{rt} = v_1 w,
}}
and all other fields presents in  \slaghc\ are set to zero\foot{The dilaton
is set to $1/3 \CR$, so that the combination $D-1/3 \CR$ vanishes.}.
The entropy function \Sen\ is defined as
\eqn\epfn{
\CE=q_I \phi^I -
2\pi \int_H d\theta d\varphi \sqrt{-\det g} \CL\Big).
}
This function depends on  free parameters $(x^I, v_1,v_2,w,\phi^I)$ of the
$SO(2,1)\times SO(3)$ ansatz~\senanz. The entropy of an extremal
black hole is obtained as an entropy of a non-extremal  black hole
in the extremal limit, when the function  \epfn\ is extremized
with respect to a free parameters
\eqn\efetr{
{\p \CE\over \p x^I} = 0, \quad {\p \CE\over \p v_1} = 0,
\quad {\p \CE\over \p v_2} = 0, \quad {\p \CE\over \p w} = 0, \quad
{\p \CE\over \p \phi^I} = 0.
}
The black hole entropy \swald\ is
given by the value of  $\CE$ at the extremum
\eqn\senfm{
S_{\rm BH} =\CE|_{\p \CE =0}.
}
The result of computation \SS\ reads
\eqn\sentr{\eqalign{
\CE = q_I\phi^I - i\pi v_1 v_2& \Big[
{1\over 4}  \Big(-{\phi^{I}\over \pi v_1}+i{p^{I}\over v_2} -
{1\over 2} x{^I}\bar w \Big)
\Big(-{\phi^{J}\over \pi v_1}+i{p^{J}\over v_2} -
{1\over 2} x{^J}\bar w \Big)\bar F{_{IJ}} +\cr
+&{\bar w \over 4}  \Big(-{\phi^{I}\over \pi v_1}+i{p^{I}\over v_2}-
{1\over 2} x{^J}\bar w \Big)F{_{I}} +
{\bar w{^2} \over 8}  F  -
\cr
-& \big({1\over v_1}-{1\over v_2}\big)\bar x{^I} F_I
+\Big( |w|^4-8 |w|^2\big({1\over v_1}+{1\over v_2}\big)
+64\big({1\over v_1}-{1\over v_2}\big)^2 \Big) F_{\hat A}
- c.c.\Big],
}}
where
\eqn\fgat{
\hat A = -4 w^2.
}
Note that the entropy function \sentr\ is invariant under the
following rescaling
\eqn\efsin{
x^I \to \la x^I, \quad w \to \la w,  \quad v_{1,2}\to {1\over \la \bar \la} v_{1,2},
\quad \phi^I \to \phi^I,
\quad q_I \to q_I, \quad p^I \to p^I,
}
since the Lagrangian \slaghc\ was derived from a superconformally invariant expression.
This means that there is a linear relation between the
extremum equations \efetr. One can switch to inhomogeneous variables
to fix this symmetry.

The above form of the entropy function does not take into account all the relevant
higher derivative corrections needed for the non-supersymmetric black hole,
as has been observed in \SS .  For example at least an $\CR ^2$ term is needed
in certain cases.  We will come back to this point in the next
section when we propose our conjecture.

To further motivate our conjecture, let us analyze the structure of the entropy function~\sentr.
First of all, compared to the topological string partition function, it depends on
one more parameter.
Indeed, using the scaling invariance of the  entropy function
(inherited from the formulation in terms of the superconformal action)
we can gauge away $w$, and identify $(X^I,W^2) \sim (x^I, \hat A)$.
However,  after that the entropy function still depends on the relative magnitude of the
variables $v_1$ and $v_2$, describing correspondingly the
radii squared of $AdS^2$ and $S_2$ factors in the black hole near horizon geometry,
and there is no such parameters in \newacr.
Therefore,
in order to match with the macroscopic  computations on the supergravity side
we need a modification
of the topological string depending on
 an additional parameter.  Moreover because of the
observations of \refs{\SS,\SSalp} this extension of topological string should
be computing additional higher derivative corrections, including extra $\CR^2$ terms.  These
observations naturally lead to our conjecture in the next section.

\newsec{A Conjecture}

In the last section we saw that we need a one parameter
extension of topological string which captures non-antiself-dual $4d$ geometries,
for higher derivative corrections for non-supersymmetric black holes.
In fact on the topological
string side
there is a natural candidate that can be used for this purpose:
a one parameter extension of the topological string that
appeared in the works of  Nekrasov \refs{\Nekrasov,\LMN,\NO,\Nstrings,\Nloc,\Nlect} on
instanton counting in Seiberg-Witten theory.
There, a function $F(X^I,\e_1,\e_2)$ was introduced.
In the special limit $-\e_2 =  \e_1 =g_{\rm top}$ this function reduces to
the ordinary topological string free energy~\ftopdf\ according to
\eqn\fnftop{
\left.F(X^I,\e_1,\e_2)\right|_{\e_1+\e_2=0} =
F_{\rm top}(X^I, g_{\rm top}), \qquad g_{\rm top}^2  = -\e_1 \e_2,
}
In order to make a connection with the supergravity ansatz \senanz\
we will need to identify the parameters as
\eqn\evid{
\e_1={16\over  |w|^2 v_1}, \quad \e_2=-{16 \over  |w|^2 v_2}.
}
This is consistent with the fact that the field theory limit $\e_{1,2}\to 0$
in the Nekrasov's approach corresponds to the flat space approximation
in the ansatz \senanz.

Since the   Nekrasov's extension of the topological string  may not be familiar,
we will first review the necessary background from \refs{\Nekrasov,\Nloc,\NY}.
Then we will be able to make a proposal about the
corresponding generalization of the OSV formula.

\subsec{Review of the Nekrasov's extension of the topological string}

The instanton corrections to the prepotential of $\CN=2$ gauge
theory can be computed by a powerful application of localization technique introduced
by Nekrasov \Nekrasov .  This localization, in the physical context gets
interpreted as turning on non-antiself-dual graviphoton background,
\eqn\rrf{
T=\e_1 dx^1 \wedge dx^2+\e_2 dx^3 \wedge dx^4 .
 }
This reproduces the $\CN =2$ prepotential by considering
the most singular term as $\epsilon_i \rightarrow 0$, which scales
as $F^{(0)}/\e_1 \e_2$.  However there is more information in the
localization computation of Nekrasov:  One can also look at the subleading
terms and identify their physical significance.  For the case of $\e_1=-\e_2$
there is a natural answer, as this gets mapped to the $\CN =2$ $F$-terms which
capture (anti)-selfdual graviphoton corrections, of the type studied in
\refs{\BCOV,\AGNT}. In fact the two can get identified using geometric engineering
of $\CN  =2$ gauge theories \refs{\KKV,\KMV} by considering,
in the type IIA setup, a local Calabi-Yau
given by ALE fibrations over some base space (e.g. $\IP^1$).  Thus Nekrasov's
gauge theory computation leads, {\it indirectly}, to a computation of topological
string amplitudes, upon the specialization $\e_1=-\e_2=g_{\rm top}$:
\eqn\fnftop{
\lim_{\e_2  \to-\e_1} F(X^I,\e_1,\e_2) =
 \sum_{g=0}^{\infty} (g_{\rm top}){^{2g-2}} F^{(g)}(X^I),
 \quad g_{\rm top}= \e_1.
}
It has been checked \refs{\IKV,\IKgeom,\HIV,\IK} using the
topological vertex formalism \refs{\AKMV,\Iqbal} that this indeed
agrees with the direct computation of topological string
amplitudes  in such backgrounds, see also
\refs{\EK,\Tachikawa,\BFFL}.

However, it is clear that there is still more to the story:  Nekrasov's computation
has more information than the topological string in such backgrounds
as it depends on an extra parameter, which is visible when $\e_1+\e_2 \not= 0$.
In fact Nekrasov's extension $F(X^I,\e_1,\e_2)$ satisfies the homogeneity condition
\eqn\homnrel{
\Big[\e_1{\p  \over \p \e_1 }+\e_2{\p  \over \p \e_2 }+
X^I{\p  \over \p X^I }\Big]F(X^I,\e_1,\e_2) =0.
}
which means that it does depend on one extra parameter compared to the topological strings.
Below we will use a shorthand notation
\eqn\shnot{
F( X,\e) \equiv F( X^I,\e_1,\e_2).
}

Even
though the exact effective field theory terms that $F( X,\e)$ computes has not been worked out, it is clear from
the derivation
that it has to do with constant, non-antiself-dual configurations of graviphoton and Riemann
curvature.  The origin of first such correction has been identified in \NY\ which we will
now review.
In general one can expand $F( X,\e) $ as follows \refs{\Nloc,\Nlect, \NY}
\eqn\znexp{
F = {1\over \e_1\e_2}F^{(0)}+{\e_1+\e_2\over \e_1\e_2}H_{1\over2}
+{(\e_1+\e_2)^2\over \e_1\e_2} G_{1}+F^{(1)}+ \CO(\e_1,\e_2).
}
Let us  discuss a geometrical meaning of the genus one terms in \znexp.
Recall a general relations
\eqn\xs{
{1\over 32 \pi^2}\int_{\CX} \Tr \CR\wedge * \CR = \chi,
\qquad  {i\over 32 \pi^2}\int_{\CX} \Tr \CR\wedge \CR ={3\over 2} \s,
}
where $\chi$ is the Euler characteristic of a Euclidean 4-manifold $\CX$ and
$\s$ is the signature.
The curvature tensor $\CR$ in \xs\ is viewed as a 2-form
$\CR^a_{\ b} = \CR^a_{\ b\m\n} dx^{\mu} \wedge dx^{\nu} $
with values in Lie algebra of $SO(4)$.
As is clear from \lvct, the ordinary topological strings compute contributions
to the effective action
of the form\foot{there is of course a similar antiholomorphic
contribution starting with $\bar F{^{(1)}}(\chi-{3\over 2}\s)$.}
\eqn\topcn{
{1\over 16 \pi^2} \int_{\CX} F^{(1)}(X) \CR_{-} \wedge  \CR_{-}
+{\rm higher \ genus} =
{1\over 2} F^{(1)} (X)\big(\chi-{3\over 2}\s\big)+{\rm higher \ genus} .
}
On the other hand, more general couplings to $\chi$ and $\s$ can be
seen in the Donaldson theory.
As was explained by Witten \Wdual, the low energy effective
action of twisted $\CN=2$ supersymmetric Yang-Mills theory on an
arbitrary four-manifold $\CX$
contains terms proportional to $\chi$  and $\sigma$.
The Donaldson invariant $D_{\xi}$ in
general has three contributions
\eqn\dinv{
D_{\xi}=Z_u +Z_+ + Z_-,
}
where $Z_{\pm}$ are Seiberg-Witten invariants defined via the
moduli space of monopoles, and~$Z_u$ is non-zero when $b_{+}(\CX)=1$ and
is given by the $u$-plane integral \MW
\eqn\zudf{
Z_u = \int_{u \rm-plane } da d\bar a  A(u)^{\chi} B(u)^{\s} e^{pu + S^2 T} \Psi.
}
As shown in \NY, the functions $A$ and $B$
are related to genus one terms in \znexp\ as
\eqn\abnek{
F^{(1)}= \ln A-{2\over 3} \ln B,\qquad
G_1={1\over 3} \ln B
}

Note that the equivariant integral of the superfield
$\Phi =\Phi^{(0)}+ \Phi^{(1)}\th+\ldots+ \Phi^{(4)}\th^4$
in the case $\CX=\IC^2$ is given by
\eqn\eqint{
\int_{\CX} d^4x \int  d^4\th \Phi ={\Phi^{(0)}(0)\over \e_1\e_2}.
}
It is also instructive to write down \NY\ the equivariant Euler number and
signature for $\IC^2$:
\eqn\eqes{
\chi(\IC^2)=\e_1\e_2, \qquad
\s(\IC^2) = {\e_1^2+\e_2^2\over 3}.
}
Let us introduce another notation:
\eqn\gff{
\tilde F^{(1)} = 4 G_1 + F^{(1)}, \qquad  G_1 ={1\over 4}(\tilde F^{(1)}- F^{(1)}).
}
Then \znexp\ can be rewritten as
\eqn\znexpn{
\e_1\e_2 F = F^{(0)}+{(\e_1+\e_2)} H_{1\over2}
+{1\over 2}\big(\chi-{3\over 2} \s\big)F^{(1)} +
{1\over 2}\big(\chi+{3\over 2}\s \big)\tilde F^{(1)} +\e_1\e_2\CO(\e_1,\e_2).
}
The term $\tilde F^{(1)} = 4 G_1 + F^{(1)}$ is not captured by the ordinary  topological string!

Extra terms are needed to obtain a correct macroscopic entropy
for non-supersym\-metric black holes   in addition to the standard
terms computed by the topological strings~\refs{\SS,\SSalp}.  In fact the particular
term needed, which is discussed in \SSalp\ reduces, upon compactification to~$4d$,
to the term of the
form $t \cdot \Tr \CR \wedge \CR$ for large $t$,
where $t$ is the overall K\"ahler moduli of the CY.
Such a correction is indeed captured by the leading behavior of $G_1 (t)$ for large $t$,
as follows from \abnek .  This gives us further confidence about
the relevance of Nekrasov's extension of topological strings for a correct
accounting of the non-supersymmetric black hole entropy.

In general, as pointed out in \HIV\ one would expect
that implementation of Nekrasov's partition function for general Calabi-Yau
will mix hypermultiplet and vector
multiplets.  The case studied in \Nekrasov\ involved the case
where there were no hypermultiplets so the question of mixing does
not arise.  In the context of the conjecture in the next section,
this would suggest that higher derivative corrections may also fix
the vevs for the hypermultiplet moduli in the context
of non-supersymmetric black holes.

We now turn to a minimal conjecture for extremal black hole entropy
which uses Nekrasov's extension of topological strings.

\subsec{Minimal $\e$-deformation }

Let us start  with a semiclassical expression \gfsem\ for
the  $\CG^{(0)}$-function
\eqn\bhpsem{\eqalign{
\CG^{(0)}={1\over 2} (P^I-X^I) (P^J-X^J) \bar
F_{IJ}^{(0)} + (P^I-X^I) F_{I}^{(0)}+ F^{(0)}, }}
where $F^{(0)}=F^{(0)}(X)$ is the Calabi-Yau
prepotential, identified with genus zero
topological string free energy, and $P^I = p^I + {i\over \pi }\phi^I$.
Our goal is to find an $\epsilon$-deformation
$\CG^{(0)} \to \CG$ of~\bhpsem, such that corresponding extremum equations
\eqn\extextreq{
{\p \i \CG \over \p \e_1 }={\p \i \CG \over \p \e_2}=
{\p \i \CG \over \p X^I}=0
}
still admit a supersymmetric attractor solution
\eqn\susgat{
\e_1={1}, \quad \e_1+\e_2 = 0, \quad X^I = P^I_{\e}= p^I + {i\over \pi }\phi^I,
}
and the extremum value of  $\i \CG$
computed using this solution is such that it describes correctly
corresponding contribution \osv\ to the  supersymmetric black hole entropy
\eqn\encorr{\eqalign{
-\i \CG_{\rm susy}\big(p^I ,\phi^I\big)=
-\i F\big(p^I + {i\over \pi }\phi^I, 256\big)=
2 \r F_{\rm top}\big(p^I + {i\over \pi }\phi^I\big).
}}

We will obtain this deformation of $\CG$-function
in two steps. First, we will
use Nekrasov's refinement  of the topological string
to deform the prepotential as
\eqn\fgenus{
F^{(0)}(X) \ \to \ F(X^I,\e_1,\e_2),
}
and at the same time, motivated from \SS, deform the  complexified magnetic charge
as\foot{When $\e_2= -\e_1 $,
this is just a rescaling of $P^I$,
while general deformation with $\e_2\not =-\e_1 $   involves
a change of the complex structure in $H^{3}(M,\IC)$.}
\eqn\psdf{
P^I \ \to \  P^I_{\e}=-\e_2 p^I + {i\over \pi} \e_1 \phi^I.
}
Second, in order to satisfy conditions \extextreq-\encorr\ after
the deformation \fgenus-\psdf, we will need to add some
compensating terms to $\CG$.
As we will see, there is some freedom in choosing these terms,
but there is  a minimal choice that does the job.

At the first step, after substituting \fgenus-\psdf\  directly into \bhpsem,
we obtain
\eqn\sezth{\eqalign{
\tilde \CG=
{1\over 2}\big(P^I_{\e}-X^I\big) \big(P^J_{\e}-X^J\big) \bar
F_{IJ}(\bar X,\bar \e) +
\big(P^I_{\e}-X^I\big) F_{I}( X,\e) +
 F( X,\e).
}}
This, however, is not the full answer, since the
derivatives of $\i\tilde \CG$ with respect to
$\e$-parameters are not zero on the supersymmetric
solution \susgat. This can be corrected at the second step, by adding to $\tilde \CG$ two
terms, proportional to $\e_1+\e_2$, so that the value \encorr\
of the potential  is not affected when $\e_1+\e_2=0$.
This leads to  the following minimal $\e$-deformation
\eqn\nwsol{\mathboxit{\eqalign{
\CG=
  {1\over 2}\big(P^I_{\e}-X^I\big) \big(P^J_{\e}-X^J\big)
\bar F_{IJ}(\bar X,\bar \e) +
\big(P^I_{\e}-X^I\big) F_{I} + F( X,\e)+ & \cr
+ {1 \over 2  } (\e_1+\e_2) \bar X{^I}F_{I}-
{1 \over 2} (\e_1+\e_2)\big(\e_1 \p_{\e_1} -\e_2 \p_{\e_2}\big) F( X,\e)&
}}}
We call \nwsol\ a minimal $\e$-deformation  because
we can also add
to \nwsol\ any terms proportional to $(\e_1+\e_2)^2$ without affecting
conditions \extextreq-\encorr:
\eqn\vnmn{
\CG \to \CG+ \CO(\e_1+\e_2)^2.}

It is straightforward to check, using the homogeneity condition \homnrel\ and  the
relations
\eqn\xphipe{
p^I = -{1\over 2 \e_2} (P_{\e}^I +\bar P{_{\e}^I} ),\qquad
\phi^I =-{i\pi\over 2 \e_1} (P_{\e}^I -\bar P{_{\e}^I} ),
}
which follow from the definition
\eqn\pedf{
P_{\e}^I = -\e_2 p^I +{i\over \pi}\e_1 \phi^I,
}
that the extremum equations \extextreq\ for
\nwsol\ indeed admit a solution \susgat,
which corresponds to a supersymmetric BPS black hole.
Moreover, in this case
\encorr\ is also satisfied.

Expression $q_I \phi^I - \pi \i \CG$ should be compared to the
entropy function \sentr.
Then our  notations are related to those of \SS\ as follows.
We identify
\eqn\evrel{
\e_1={16\over |w|^2 v_1}, \quad \e_2=-{16 \over |w|^2 v_2}.
}
The supersymmetric attractor equations of \SS\ read
$
p^I =-{i\over 4}v_2(\bar w  x^I-w \bar x{^I}),
$
while in our conventions the supersummetric case is $p^I = \r X^I$. Therefore,
\eqn\xXrel{
X^I=- {i\over 2}\bar   w  x^I,\qquad x^I = {2 i  \over \bar w   } X^I.
}
We also set in this case
\eqn\wgauge{
w\bar w = 16,\quad v_1=v_2=1.
}
%

\subsec{Putting it all together}

Now we are ready to make a proposal about the
extremal black holes entropy. We want to write down a generalization
of the semiclassical expression for the extremal black hole
degeneracy from section 5, that would reduce to the OSV
formula \osvcn\ for the supersymmetric charge vector $(p^I, q_I)$.
The expression \nwsol\ for the deformed  black hole potential
provides a natural way to do this,
and allows to treat supersymmetric and non-supersymmetric cases simultaneously.

We  introduce a  function $\CG=\CG(p,\phi; X,\e)$ defined by
\eqn\gfundfn{\eqalign{
\CG=
{1\over 2}\big(P^I_{\e}-X^I\big) \big(P^J_{\e}-X^J\big)
\bar F_{IJ}(\bar X,\bar \e) +
\big(P^I_{\e}-X^I\big) F_{I}( X,\e) + F( X,\e)+ ~~~~~~ & \cr
+ {1 \over 2  } (\e_1+\e_2) \bar X{^I}F_{I}( X,\e)-
{1 \over 2} (\e_1+\e_2)\big(\e_1 \p_{\e_1} -\e_2 \p_{\e_2}\big) F( X,\e)+
\CO(\e_1+\e_2)^2&,
}}
where $\CO(\e_1+\e_2)^2$ denotes an ambiguity that
cannot be fixed just by requiring that $\i \CG$ gives
correct description of the supersymmetric black holes.
In the minimal deformation case we set $\CO(\e_1+\e_2)^2=0$.
In general, there are two types of solutions
to the extremum equations
\eqn\veextr{
{\p \over \p X^I }\i \CG = {\p \over \p \e_i }\i \CG = 0,
}
the supersymmetric one \susgat\ $X^I = p^I + {i\over \pi} \phi^I$,
and non-supersymmetric ones (all other).
Let us denote the functions  obtained by substituting
these non-supersymmetric solutions $X^I=X^I(p,\phi), \ \e_{1,2}=\e_{1,2}(p,\phi)$
into \gfundfn\ as
$ \CG(p^I,\phi^I)$.
For supersymmetric solution
$\CG_{\rm susy}(p^I,\phi^I) =F(p^I+{i\over \pi}\phi^I)$.
We conjecture the following relation for the extremal black hole degeneracy
\eqn\genrl{\mathboxit{
\Omega_{\rm extrm}(p^I,q_I) = \int d\phi^I e^{q_I \phi^I} \Big(
\Big|e^{{i \pi\over 2} F(p^I+{i\over \pi}\phi^I)}\Big|^2+
\sum_{\rm n-susy} \Big|e^{ {i\pi \over 2} \CG(p^I,\phi^I)}\Big|^2
\Big),
}}
which is expected to be valid asymptotically in the
limit of large charges.
The sum in \genrl\ runs over all non-supersymmetric solutions to the
extremum equations \veextr. However, it is expected
that for a given set of charges $(p^I,q_I)$ only one
solution (supersymmetric or non-supersymmetric, depending on
the value of the discriminant) dominates, and contributions from all other solutions,
including the ones with non-positive Hessian, are exponentially suppressed.

As noted before, it is expected that for general non-toric Calabi-Yau
compactifications, which lead to
hypermultiplets, the analog of Nekrasov's partition function would mix
hypermultiplets with vector multiplets and therefore will fix
their values at the horizon.  This would be interesting
to develop further.

\newsec{Conclusions and Further Issues }

We studied the  black hole potential
describing  extremal black hole solutions in $\CN=2$ supergravity
and found a new formulation of the semi-classical
attractor equations, utilizing
homogeneous coordinates on the Calabi-Yau moduli space.
This allowed us to solve
the inverse problem (that is, express the  black hole
charges in terms of the attractor  Calabi-Yau moduli)
completely in the one-modulus Calabi-Yau case. We found three
non-supersymmetric solutions in addition to the supersymmetric one.
In the higher dimensional case we found a bound
$
\#_{\rm n-susy}\leq 2^{n_V+1}-1
$
on the possible  number of non-supersymmetric solutions
to the inverse problem.

We then investigated a generalization of the
attractor equations and OSV formula in
the case when other corrections are turned on.
We conjectured that corresponding corrected
extremal black hole entropy needs an additional ingredient:
the Nekrasov's extension of the topological
string free energy  $ F(X^I,\e_1,\e_2)$.  We related
this to the black hole entropy using a minimal deformation
conjecture given in \nwsol,\genrl,
that reduces to $F_{\rm top} (X^I + {i\over \pi} \phi^I)$ for the choice
of the black hole charges that support a supersymmetric
solution.   We were unable to fix the $\CO \big(\e_1+ \e_2 \big)^2$
ambiguity in \gfundfn, though it could be that the
minimal conjecture is correct.

 One important open
question is how to test our conjecture.  One
possible test may be using the local Calabi-Yau geometry for which
Nekrasov's partition function is known.  Another important
question is to find out what is exactly computed by
Nekrasov's partition function\foot{for example, in the $AdS_2\times S^2 $ setup
of \SS, the $\e$-parameters corresponding to the radii of $AdS_2$ and $S^2 $
factors were real, but from the topological string  viewpoints it is natural
to consider a complexification of $\e_{1,2}$.
This suggests that there should exist
corresponding deformation of the $AdS_2\times S^2 $ near horizon geometry.
} and how to extend it to the
case where there are both hypermultiplets and vector multiplets.
Clearly there is a long road ahead.  We hope to have provided
strong evidence that Nekrasov's extension of topological string
is a key ingredient in a deeper understanding of non-supersymmetric
black holes.


\break

\centerline{\bf Acknowledgments}

We would like to thank R. Dijkgraaf, A. Neitzke, M. Rocek and A. Sen  for
valuable discussions.
We would also like to thank the Stony Brook physics department and 4th
Simons Workshop in Mathematics and Physics for providing a
stimulating environment where part of this work was done.
CV thanks the CTP at MIT for hospitality during his sabbatical leave.
KS thanks the hospitality of the theory group at Caltech during the final stage
of this work.
The work of the authors was supported in part by
NSF grants PHY-0244821 and DMS-0244464.
This research of KS was  supported in part by
{\cyr RFFI} grant 07-02-00645, and  by the grant {\cyr NSh}-8004.2006.2
 for scientific schools.
%


\appendix{A}{The Diagonal Torus Example}

Consider the case \Moore\ when $M=T^6$ is
the so-called diagonal torus:
\eqn\dt{
M= \Sigma_{\tau} \times \Sigma_{\tau}  \times \Sigma_{\tau},
}
where  $\Sigma_{\tau} = \IC/(\IZ+ \tau \IZ)$ is the elliptic
curve with modular parameter $\tau$.
Let us introduce complex coordinates  $dz^i=dx^i+\tau dy^i, i=1,2,3$
on each $\Sigma_{\tau}$.
As in \BD\ can label the relevant 3-cycles of $M$ according to their
mirror branes in IIA picture:
\eqn\dbr{\eqalign{
D0 \to &-dy^1 dy^2 dy^3 \cr
D2 \to &dy^1 dy^2 dx^3 + dy^1 dx^2 dy^3 + dx^1 dy^2 dy^3 \cr
D4 \to &dx^1 dx^2 dy^3 + dx^1 dy^2 dx^3 + dy^1 dx^2 dx^3 \cr
D6 \to &-dx^1 dx^2 dx^3
}}
The intersection matrix of these 3-cycles is
\eqn\imat{
\pmatrix{
0 & 0 & 0 & -1\cr
0 & 0 & 3 & 0 \cr
0 & -3 & 0 & 0 \cr
1 & 0  & 0 & 0
}.
}
We denote the brane charge vector as $(D0,D2,D4,D6)=(u,q,p,v)$.
Then
\eqn\wtt{
\CW=u + 3 q\tau - 3 p\tau^2-v\tau^3.
}
The black hole potential is
\eqn\vbh{
V_{\rm BH} = e^K\big(|\CW|^2+g^{\tau \bar \tau}|\p \CW + \CW \p K |^2\big)
}
where
\eqn\wts{
K \sim \log(\i \tau)^3, \qquad
g^{\tau \bar \tau} = {3\over 4 (\i \tau)^2}.
}
Therefore, we have
\eqn\vbhe{
V_{\rm BH} ={8\over (\i \tau )^3}
\Big( \big| u+3 q\tau -3 p \tau^2 -v \tau^3 \big|^2 +
3 \big|2 i \i \tau (q - 2 p \tau -v \tau^2)
- u - 3 q\tau +3 p \tau^2 +v \tau^3 \big|^2
\Big).
}
%

\subsec{Solution of the inverse problem}

Let us decompose $\tau$ into the real and imaginary parts
\eqn\tir{
\tau=\tau_1+i \tau_2,
}
and introduce new variables $\a,\b,\g$
that are real linear combination of the charges
\eqn\abg{\eqalign{
\a=&\CW\big|_{\tau_2=0}=u+3q\tau_1-3p\tau_1^2-v\tau_1^3,\cr
\b=&{1\over 3} {\p \CW\over \p \tau}\Big|_{\tau_2=0}=q-2p\tau_1-v\tau_1^2,\cr
\g=&-{1\over 6} {\p^2 \CW\over \p \tau^2}\Big|_{\tau_2=0} =p+v\tau_1. }}
Using \abg, we can  rewrite the superpotential \wtt\ as
\eqn\wxy{
\CW = \a+ 3 i \b \tau_2+ 3 \g \tau_2^2+iv\tau_2^3.
}
Then \wcnr\ gives
\eqn\leqn{\eqalign{
\a+3 \g \tau_2^2=&\om_1, \cr
3 \b \tau_2+v\tau_2^3= &\om_2,
}}
where $\om=\om_1+i\om_2$. The black hole
potential \vdf\ in new variables is given by
\eqn\vbhab{
 V_{\rm BH}={32\over \tau_2^3}\big(
\a^2+3\b^2\tau_2^2+3\g^2\tau_2^4+v^2\tau_2^6 \big)
}
The extremum equations
${\p V_{\rm BH}\over \p \tau_1}={\p V_{\rm BH}\over \p \tau_2}=0$
take the form:
\eqn\exts{\eqalign{
\a\b-2\b\g \tau_2^2+v\g \tau_2^4=&0, \cr
-\a^2-\b^2 \tau_2^2+\g^2 \tau_2^4+v^2 \tau_2^6=&0,
}}
If we express $\a$ and $v$ in terms of
$\b$ and $\g$ using \leqn\
\eqn\avbg{\eqalign{
\a=\om_1-3\g\tau_2^2, \qquad
v={\om_2-3\b\tau_2\over\tau_2^3},
}}
assuming $\tau_2\not =0$,
the first equation in \exts\ gives
\eqn\bsol{
\b= {\g \om_2 \tau_2 \over 8 \tau_2^2 \g-\om_1}.
}
Here we also assumed that $ 8 \tau_2^2 \g\not =\om_1$. We will discuss this
special case later.
The second equation in \exts\ then  takes the form
\eqn\sefm{
\big(4\tau_2^2\g-\om_1\big)\big(128 \tau_2^6 \g^3 - 96 \tau_2^4\g^2 \om_1 +
18 \tau_2^2 \g \om_1^2  -6 \tau_2^2 \g \om_2^2 + \om_1 \om_2^2- \om_1^3 \big)
=0.
}
We immediately see that $\g={\om_1\over 4\tau_2^2}$, and therefore
\eqn\stssl{
\a={\om_1\over 4},
\qquad \b={\om_2\over 4\tau_2},
\qquad \g={\om_1\over 4\tau_2^2},
\qquad v={\om_2\over 4\tau_2^3},
}
gives a solution to \exts. In fact, it describes  a supersymmetric branch of the
extremum equations \extrm.
The cubic equation for $\g$ in \sefm\ has
three non-susy solutions that can be described by the formula:
\eqn\nsgr{
\g={2 \r (\om)+ |\om| \big({|\om|/ \om}\big)^{1/ 3}+
|\om| \big({|\om|/ \om}\big)^{-1/3} \over 8 \tau_2^2},
}
where one can choose any of three cubic root branches.
It is obvious that all solutions \nsgr\ are real.
Correspondingly, in this case
\eqn\nsbs{\eqalign{
\a=& {1\over 4}\r(\om) -
{3\over 8}|\om| \big({|\om|/ \om}\big)^{1/ 3}-
{3\over 8} |\om| \big({|\om|/ \om}\big)^{-1/3},    \cr
\b=&{ \i (\om) \over 8 \tau_2} \cdot
{2 \r (\om)+ |\om| \big({|\om|/ \om}\big)^{1/ 3}+
|\om| \big({|\om|/ \om}\big)^{-1/3}  \over
 \r (\om)+ |\om| \big({|\om|/ \om}\big)^{1/ 3}+
|\om| \big({|\om|/ \om}\big)^{-1/3} }, \cr
\g=&{2 \r (\om)+ |\om| \big({|\om|/ \om}\big)^{1/ 3}+
|\om| \big({|\om|/ \om}\big)^{-1/3} \over 8 \tau_2^2}, \cr
v=& { \i (\om) \over 8 \tau_2^3} \cdot
{2 \r (\om)+ 5|\om| \big({|\om|/ \om}\big)^{1/ 3}+
5|\om| \big({|\om|/ \om}\big)^{-1/3}  \over
 \r (\om)+ |\om| \big({|\om|/ \om}\big)^{1/ 3}+
|\om| \big({|\om|/ \om}\big)^{-1/3} } .
}}
It is instructive to compute the values of the
black hole potential \vbhab\ at the three non-supersymmetric
extremal points \nsbs. Using the second equation in \exts, we
obtain
\eqn\vbhns{
V_{\rm BH} = {64 \over \tau_2}\big(
\b^2+2\g^2\tau_2^2+v^2\tau_2^4
\big).
}
If we apply \avbg, after some algebra we find
\eqn\vsmpl{\eqalign{
\b^2+2\g^2\tau_2^2+v^2\tau_2^4=
{128 \tau_2^8 \g^4 \! -\! 32\tau_2^6\g^3\om_1+2\tau_2^4\g^2\om_1^2+26\tau_2^4\g^2\om_2^2
-10\tau_2^2\g\om_1\om_2^2+\om_1^2\om_2^2 \over
\tau_2^2(8\tau_2^2\g-\om_1)^2}=\cr
={\om_1^2+\om_2^2\over 2 \tau_2^2}+
{\tau_2^2\g+\om_1/2\over \tau_2^2(8\tau_2^2\g-\om_1)^2}
\big(128 \tau_2^6 \g^3 - 96 \tau_2^4\g^2 \om_1 +
18 \tau_2^2 \g \om_1^2  -6 \tau_2^2 \g \om_2^2 + \om_1 \om_2^2- \om_1^3 \big).
}}
The last term in the second line vanishes at the non-supersymmetric
extremum point due to \sefm, and we get a simple formula for the
potential
\eqn\vbhnse{
V_{\rm BH}^{\rm n-susy} = 32  {|\om|^2 \over \tau_2^3} .
}
Notice that the value of the potential  is the same for all three points \nsbs.
At the supersymmetric extremum point \stssl\ we have
\eqn\vbhse{
V_{\rm BH}^{\rm susy} = 8  {|\om|^2 \over \tau_2^3},
}
so that, as in \BFM
\eqn\vsnsrel{
V_{\rm BH}^{\rm n-susy}=4V_{\rm BH}^{\rm susy}.
}
Note that this relation is written in terms of Calabi-Yau moduli
rather then in terms of the black hole charges.

As we will see in a moment, all three  non-supersymmetric extremum
points  provide a minimum of the  black hole potential.
In order to show this,
let us look at the Hessian
\eqn\ness{\eqalign{
Hess(V_{\rm BH})=\pmatrix{{\p^2 V_{\rm BH}\over \p \tau_1^2 } &
{\p^2 V_{\rm BH}\over \p \tau_1 \p \tau_2} \cr
{\p^2 V_{\rm BH}\over \p \tau_2 \p \tau_1} &
{\p^2 V_{\rm BH}\over \p \tau_2^2 } }.
}}
Straightforward computation  gives
\eqn\hsns{
Hess(V_{\rm BH})=
{192\over \tau_2^3}\pmatrix{
3\b^2-2 \a\g+(4\g^2-2\b v)\tau_2^2+v^2\tau_2^4 & 4\g\tau_2(-\b+v\tau_2^2)\cr
4\g\tau_2(-\b+v\tau_2^2) & -\b^2+2\g^2\tau_2^2+3 v^2\tau_2^4
}.
}
At the non-supersymmetic extremal point \nsbs,
using \avbg\ and \sefm, we obtain the following expression
\eqn\hsnxt{\eqalign{
M&={\tau_2^3\over 96} Hess(V_{\rm BH})=\cr
\! =\! &\pmatrix{\! \!
{96 \tau_2^4 \g^2 (2\om_1^2+\om_2^2)-8\tau_2^2\g\om_1(6\om_1^2+\om_2^2)
+3\om_1^4-\om_1^2\om_2^2\over
\tau_2^2(8\tau_2^2\g-\om_1)^2} & \! \!
{8\g(4\tau_2^2\g-\om_1)\om_2 \over 8\tau_2^2\g-\om_1}\cr \! \!
{8\g(4\tau_2^2\g-\om_1)\om_2 \over 8\tau_2^2\g-\om_1} & \! \! \!
{32 \tau_2^4 \g^2 (2\om_1^2+5\om_2^2)-8\tau_2^2\g\om_1(2\om_1^2+7\om_2^2)
+\om_1^4+5\om_1^2\om_2^2\over
\tau_2^2(8\tau_2^2\g-\om_1)^2 \! \! }
}
}}
The eigenvalues $h_{1,2}$ of the matrix \hsnxt\ are
solutions to the equation
\eqn\hesev{\eqalign{
0=\det\left\|M-\pmatrix{h & 0 \cr 0& h}\right\|=
h^2-4{|\om|^2\over \tau_2^2} h+3{|\om|^4\over \tau_2^4}&- \cr
\! -{8 \om_2^2(4\tau_2^2\g \! -\! \om_1)
(16 \tau_2^4 \g^2 \! + \! 4 \tau_2^2 \g \om_1 \! -\! \om_1^2)\over
\tau_2^4 (8\tau_2^2\g-\om_1)^4}
\big(128 \tau_2^6 \g^3\! -\! 96 \tau_2^4\g^2 \om_1 +&
6 \tau_2^2 \g (3 \om_1^2 \! - \! \om_2^2) \! + \! \om_1 \om_2^2 \! -\! \om_1^3 \big)
}}
The last line vanishes because of the extremum equation \sefm,
and we get
\eqn\hev{
h^2-4{|\om|^2\over \tau_2^2} h+3{|\om|^4\over \tau_2^4}=0.
}
Therefore, the eigenvalues of the matrix \hsnxt\
\eqn\evexp{\eqalign{
h_1 =& {|\om|^2\over \tau_2^2} \geq 0 \cr
h_2 =& 3{|\om|^2\over \tau_2^2}\geq 0
}}
are always non-negative.
Since $ \tau_2>0$, this means that the eigenvalues of the
Hessian \hsns\ are also positive if $\om\not=0$,
and thus the non-supersymmetric
extremum points minimize the potential.

\subsec{Solution of the direct problem}

The black hole potential \vbhe\  is given by
\eqn\vbhab{\eqalign{ V_{\rm BH} =& {4\over \tau_2^3} \big( u^2 + 6
q u \tau_1 + 9 q^2 \tau_1^2 - 6 p u\tau_1^2 - 18 p q \tau_1^3 -
    2 u v \tau_1^3 + 9 p^2 \tau_1^4 -  \cr & - 6 q v \tau_1^4 + 6 p v \tau_1^5 +
    v^2 \tau_1^6
    + 3 q^2 \tau_2^2  - 12 p q \tau_1 \tau_2^2 + 12 p^2 \tau_1^2 \tau_2^2 -
    \cr &  - 6 q v \tau_1^2 \tau_2^2 +  12 p v \tau_1^3 \tau_2^2 +
    3 v^2 \tau_1^4 \tau_2^2 +
    3 p^2 \tau_2^4 + 6 p v \tau_1 \tau_2^4 + 3 v^2 \tau_1^2 \tau_2^4 + v^2 \tau_2^6
\big).
}}
Straightforward calculation gives
\eqn\vda{ {\p V_{\rm BH} \over \p \tau_1}= {24\over \tau_2^3}
\Big(\! (q-2p\tau_1-v\tau_1^2)(u+3q\tau_1-3p\tau_1^2-v\tau_1^3)-
2(p+v\tau_1)(q-2p\tau_1-v\tau_1^2)\tau_2^2 + (p+v\tau_1)v \tau_2^4
\Big) }
and
\eqn\vdb{
{\p V_{\rm BH} \over \p \tau_2}={12\over \tau_2^4} \Big(
-(u+3q\tau_1-3p\tau_1^2-v\tau_1^3)^2-(q-2p\tau_1-v\tau_1^2)^2
\tau_2^2+ (p+v\tau_1)^2\tau_2^4+v^2 \tau_2^6 \Big).
}
The extremal points are solutions to the equations ${\p V_{\rm BH}
\over \p \tau_1}={\p V_{\rm BH} \over \p \tau_2}=0$. From \vda\ we
find that for a generic set of charges (assuming $v\g\not =0$)
\eqn\bsol{
\tau_2^2={\b\g \pm \sqrt{\b\g(\b\g-v \a)} \over v \g},
}
where
\eqn\abg{\eqalign{
\a=&u+3q\tau_1-3p\tau_1^2-v\tau_1^3,\cr
\b=&q-2p\tau_1-v\tau_1^2,\cr
\g=&p+v\tau_1. }}
If we plug  \bsol\ into \vdb, we obtain
\eqn\fctr{
\g \sqrt{\b\g-v \a} \big(
\b\sqrt{\b\g} (v^2\a-3v \b\g-2\g^3) \mp
\g \sqrt{\b\g-v \a} (3v\b^2+v\a\g+2\b\g^2)
\big)=0.
}
Let us look at the solution $\b\g-v \a=0$ first.
Due to \abg\ this is equivalent to
\eqn\sasol{
\tau_1={ p q - u v\over 2(p^2  + q v)}
}
Then \bsol\ gives, assuming $\tau_2>0$
\eqn\sbsol{
\tau_2={\sqrt{-\CD}\over
2(p^2  + q v)} }
where
\eqn\ddf{\CD=-\big(3 p^2 q^2+4 p^3 u+4 q^3 v+6 p q u v-u^2 v^2\big).
}
This is the supersymmetric solution obtained in \Moore.
Note that there is no such solution if the discriminant
\ddf\ is positive: $\CD > 0$.

The non-supersymmetric solution will emerge from the
second branch:
\eqn\nssl{
\b\sqrt{\b\g} (v^2\a-3v \b\g-2\g^3) =\pm
\g \sqrt{\b\g-v \a} (3v\b^2+v\a\g+2\b\g^2)
}
Without loss of generality we can take the square of this equation.
Then, after plugging in \abg\ we find massive cancellations,
and obtain the following cubic
equation
\eqn\cx{\eqalign{
\big(2 p^6+6 p^4 q v+3 p^2 q^2 v^2-4 p^3 u v^2-2 q^3 v^3-6 p q u v^3+u^2 v^4\big)& \tau_1^3-\cr
-3 (p^5 q+5 p^3 q^2 v+3 p^4 u v+5 p q^3 v^2+4 p^2 q u v^2-q^2 u v^3-p u^2 v^3\big)&\tau_1^2-\cr
-3 \big(p^4 q^2+2 p^5 u+2 p^3 q u v-2 q^4 v^2-2p q^2 u v^2-p^2 u^2 v^2\big)& \tau_1\ +\cr
+\big(2 p^3 q^3+3 p^4 q u+3 p q^4 v+6 p^2 q^2 u v+p^3 u^2 v+q^3 u v^2 \big) & \ =\ 0.
}}
The discriminant of this equation  is equal to
\eqn\dscc{\Delta=729\CD^3 \big(p^2+q v \big)^6
\big(2 p^6+6 p^4 q v+3 p^2 q^2 v^2-4 p^3 u v^2-2 q^3 v^3-6 p q u v^3+u^2 v^4\big)^2.
}
Only one solution of this equation can be real, if $\CD>0$, which
implies $\Delta>0$, but this is
exactly what we are looking for. It is given by
\eqn\xsns{\eqalign{
\tau_1=&{1\over (2 (p^2+q v)^3+v^{2} \CD)}
\Big( (p^2+q v )^2(pq-uv) -v p \CD  - \cr &-
{2^{1/3}  (p^2+q v )^3 \CD \over  \big( v    (2 p^3+3 p q v-u v^2 ) \CD^2+
  (2  (p^2+q v )^3 +v^{2 }\CD ) \CD \sqrt{  \CD  }\big)^{1/3}}  + \cr &+
 {p^2+q v \over 2^{1/3} } \big( v   (2 p^3+3 p q v-u v^2 ) \CD^2+
(2  (p^2+q v )^3 +v^{2 }\CD  ) \CD\sqrt{  \CD  } \big)^{1/3}
\Big).
}}
Corresponding expression for $\tau_2$ is obtained by
substituting \xsns\ into \bsol.

\appendix{B}{Cubic equation}

Consider a general cubic equation of the form
\eqn\cbc{
 ax^3 + 3 b x^2 - 3 c x- d=0.
}
The discriminant of this equation is
\eqn\cds{
\Delta=-(3b^2c^2+4c^3a+4b^3d+6abcd-a^2d^2).
}
The solutions are given by
\eqn\so{\eqalign{
x_1=
-{b \over a}+{2^{1/3} (b^2+ a c) \over
a \big(a^2 d  -3 abc-2 b^3  + a\sqrt{\Delta} \big)^{1/3}
}+
{\big( a^2 d  -3 abc-2 b^3  + a\sqrt{\Delta}  \big)^{1/3}\over  2^{1/3} a},
}}
\eqn\st{\eqalign{
x_2=
-{b \over a}-{2^{1/3} (1+i\sqrt{3})(b^2+ a c) \over
2 a \big( a^2 d  -3 abc-2 b^3  + a\sqrt{\Delta}
\big)^{1/3}
}- {(1-i\sqrt{3})\over  2^{1/3} 2 a}
\big( a^2 d  -3 abc-2 b^3  + a\sqrt{\Delta}
\big)^{1/3},
}}
\eqn\str{\eqalign{
x_3=
-{b \over a}-{2^{1/3} (1-i\sqrt{3})(b^2- a c) \over 2 a \big(
a^2 d  -3 abc-2 b^3  + a\sqrt{\Delta}
\big)^{1/3}
}- {(1+i\sqrt{3})\over  2^{1/3} 2 a}
\big( a^2 d  -3 abc-2 b^3  + a\sqrt{\Delta}
\big)^{1/3}
}}
We are interested in the case $\Delta >0$, when there is
one real root and a pair of complex conjugate roots.

\listrefs
\end